\documentclass[aps,prl, showpacs,groupedaddress,preprint,floatfix,longbibliography]{revtex4-1}

\usepackage{amsmath}
\usepackage{graphicx}
\usepackage[usenames,dvipsnames]{color}
\pdfoutput=1

\usepackage{color}
\usepackage{url}
\usepackage[colorlinks]{hyperref}
\hypersetup{%
        plainpages=true,
        breaklinks=true,
        hypertexnames=false,
        pageanchor=true,
        colorlinks=true,
        linkcolor={blue},
        citecolor={blue},
        urlcolor={black}, 
        anchorcolor={black}
      }

\renewcommand{\eqref}[1]{\mbox{Eq.~(\ref{#1})}}

\begin{document}


\title{A Single Photon Can Simultaneously Excite Two or More Atoms}

\author{Luigi Garziano$^{1,2}$}

\author{Vincenzo Macr\`{i}$^1$}
\author{Roberto Stassi$^{1,2}$}
\author{Omar Di Stefano$^1$}
\author{Franco Nori$^{2,3}$}
\author{Salvatore Savasta$^{1,2}$}

\affiliation{$^1$Dipartimento di Fisica e di Scienze della Terra, Universit\`{a} di Messina, I-98166 Messina, Italy,}
\affiliation{$^2$CEMS, RIKEN, Saitama 351-0198, Japan,}
\affiliation{$^3$Physics Department, The University of Michigan, Ann Arbor, Michigan 48109-1040, USA}



\begin{abstract}
We consider two separate atoms interacting with a single-mode optical resonator. When the frequency of the resonator field is twice the atomic transition frequency, we show that there 
exists a resonant  coupling between \textit{one} photon and \textit{two} atoms,  via intermediate virtual states connected by counter-rotating processes. If the resonator is prepared in its one-photon state, the photon can be jointly absorbed by the two atoms in their ground state which will both reach their excited state with probability close to one. Like ordinary quantum Rabi oscillations, this process is coherent and reversible, so that two atoms in their excited state will undergo a downward transition jointly emitting a single cavity photon. 
This joint absorption and emission processes can also occur with three atoms. The parameters used to investigate this process correspond to experimentally demonstrated values in circuit quantum electrodynamics systems.


\end{abstract}


\maketitle




Multiphoton excitation and emission processes were predicted in 1931 by Maria G{\"{o}ppert-Mayer  in her doctoral dissertation on the theory of two-photon quantum transitions \cite{Goeppert-Mayer1931}. Two-photon absorption consists in the simultaneous absorption of two photons of identical or different frequencies by an atom or a molecule.
Two-photon excitation is now a powerful spectroscopic and diagnostic tool \cite{Denk1990,So2000}. 
One may wonder if the reverse phenomenon, {i.e.}, joint multiatom emission of one photon or  multiatom excitation with a single photon, is ever possible.
We show that these processes, not only can be enabled by the strong correlation between the states of the atoms and those of the field occurring in cavity quantum electrodynamics (QED) \cite{Haroche2006}, but they can even take place with probability approaching one.

Cavity QED investigates the interaction of confined electromagnetic field modes with natural or artificial atoms  under conditions where the quantum nature of light affects the system dynamics \cite{Haroche2013,Haroche2013a}.
 A high degree of manipulation and control of quantum systems can be reached in the strong-coupling regime, where the atom-field coupling rate is dominant with respect to the loss and decoherence rates.
 This paves the way for
many interesting physical applications \cite{Deleglise2008, Hofheinz2009,Haroche2013a, Vlastakis2013}.
Cavity QED is also very promising for the realization of  quantum gates \cite{Rauschenbeutel1999, Zheng2000, You2006} and quantum networks for quantum computational tasks \cite{Felinto2006,Kimble2008,Ritter2012}.
Many of the proposed concepts, pioneered with flying atoms,
have been adapted and further developed using superconducting
artificial atoms in the electromagnetic field of microwave
resonators, giving rise
to the rapidly growing  field of circuit QED, which is very
promising for future quantum technologies \cite{You2003,Blais2004,Wallraff2004,Chiorescu2004,You2006, Hofheinz2009,Vlastakis2013}.
In these systems, coupling rates between an individual qubit and a single electromagnetic mode of the order of $10 \%$ of the unperturbed frequency
 of the bare subsystems have been experimentally reached \cite{Niemczyk2010, Forn-Diaz2010, Fedorov2010,Forn-Diaz2015}. Such a coupling rate is significantly higher than that obtained using natural atoms. 
 Such an ultrastrong coupling (USC) opens the door to the study of the physics of virtual processes which do not conserve the number of excitations governed by the counter-rotating terms in the interaction Hamiltonian \cite{DeLiberato2009,Ai2010,Cao2010,Cao2011,Stassi2013,Ridolfo2013,Garziano2013,Garziano2014,Huang2014,Zhao2015}.
Recently, it has been shown that these excitation-number-nonconserving processes enable higher order atom-field  resonant transitions, making possible  coherent and reversible multiphoton exchanges between the qubit and the resonator \cite{Zhu2013,Law2015,Garziano2015}.

Here we examine  a quantum system constituted by two two-level atoms coupled to a single-mode resonator in the regime where the field-atom detuning $\Delta =\omega_{\rm c} - \omega_{\rm q}$ is large ($\omega_{\rm c}$ and $\omega_{\rm q}$ are the resonance frequency of the cavity mode  and the qubit transition frequency). 
We investigate the situation where the two qubits are initially in their ground state and one photon is present in the resonator, corresponding to the initial state $|g,g,1 \rangle$.  We find that, if $\omega_{\rm c} \approx 2 \omega_{\rm q}$, a single cavity photon  is able to excite simultaneously two independent atoms. During this process no parametric down-conversion,  splitting the initial photon into observable pairs of photons at frequency $\omega_{\rm c}/2$, occurs. The cavity photon is directly and jointly absorbed by the two atoms.
As shown in Fig.~1, the initial state $|g,g,1 \rangle$ goes to  virtual
intermediate states that do not conserve the energy, but comes back to the real final state
$|e,e,0 \rangle$ that does conserve energy.
 If $\omega_{\rm c} \approx 3 \omega_{\rm q}$ the simultaneous excitation of three atoms:  $|g,g,g,1 \rangle \to |e,e,e,0 \rangle$ is also possible. If the coupling is sufficiently strong, even a higher number of atoms can be excited with a single photon. Owing to optical selection rules, the two-atom process requires  parity-symmetry breaking of the atomic potentials, which can be easily  achieved in superconducting artificial atoms \cite{Liu2005,Gross2008,Zhu2013}. On the contrary, the three-atom  process does not need broken symmetry.

The Hamiltonian describing the system consisting of a single cavity mode interacting with two or more identical qubits with possible symmetry-broken potentials  is given by \cite{Niemczyk2010,Garziano2014}
\begin{equation}
\hat H_0 =  \hat H_{\rm q} + \hat H_{\rm c} + \lambda  \hat X
 \sum_i
 (\cos \theta\,  \hat \sigma^{(i)}_x + \sin \theta\, \hat \sigma^{(i)}_z )\, ,
\label{HH}
\end{equation}
where  $\hat H_{\rm q} = ({\omega_{\rm q}}/{2}) \sum_i  \hat \sigma^{(i)}_z $ and $ \hat H_{\rm c} =\omega_{\rm c} \hat a^\dag \hat a$, describe the qubit and cavity Hamiltonians in the absence of interaction, $\hat X =  \hat a + \hat a^\dag$, $\hat \sigma^{(i)}_x$ and $\hat \sigma^{(i)}_z$
are Pauli operators for the $i$th qubit, and $\lambda$ is the coupling rate of each qubit to the cavity mode. For $\theta= 0$ parity is conserved. For flux qubits, this angle, as well as the transition frequency $\omega_{\rm q}$, can be continuously tuned by changing the external flux bias \cite{Liu2005,Niemczyk2010}. For the sake of simplicity, Eq.~(\ref{HH}) describes identical qubits, but this is not an essential point.
In contrast to the Jaynes-Cummings (JC) model, the Hamiltonian in Eq.~(\ref{HH}) explicitly contains counter-rotating
 terms of the form $\hat \sigma^{(i)}_+ \hat a^\dag$, $\hat \sigma^{(i)}_- \hat a$, $\hat \sigma^{(i)}_z \hat a^\dag$, and $\hat \sigma^{(i)}_z \hat a$. The first (second) term creates (destroys) two excitations while the third (fourth) term creates (destroy) one excitation.
The presence of counter-rotating terms in the interaction Hamiltonian enables four different paths which, starting from the initial state $|g,g,1 \rangle$,  reach the final state $|e,e,0 \rangle$ (see Supplemental Fig.~S1 \cite{SuppMat}). Each path includes three virtual transitions involving out-of-resonance intermediate states. Fig.~1 only displays  the process that gives the main contribution to the effective coupling between the bare states $|g,g,1 \rangle$ and  $|e,e,0 \rangle$. Higher-order processes, depending on the atom-field interaction strength, can also contribute. By applying standard third-order perturbation theory, we obtain the following effective coupling rate, $\Omega_{\rm eff}/\omega_{\rm q}  \equiv  (8/3)(\lambda/\omega_{\rm q})^3 \sin \theta \cos^2{\theta}$. The analytical derivation of the  effective coupling rate as a function of $ \lambda/\omega_{\rm q} $ is presented in the Supplemental Material \cite{SuppMat}. Already at a coupling rate $\lambda/ \omega_{\rm q} = 0.1$, an effective (two qubits)-(one photon) coupling rate $\Omega_{\rm eff}/\omega_{\rm q} \sim 10^{-3}$ can be obtained. 
\begin{figure}[!ht]
	\centering
	\includegraphics[scale=0.20]{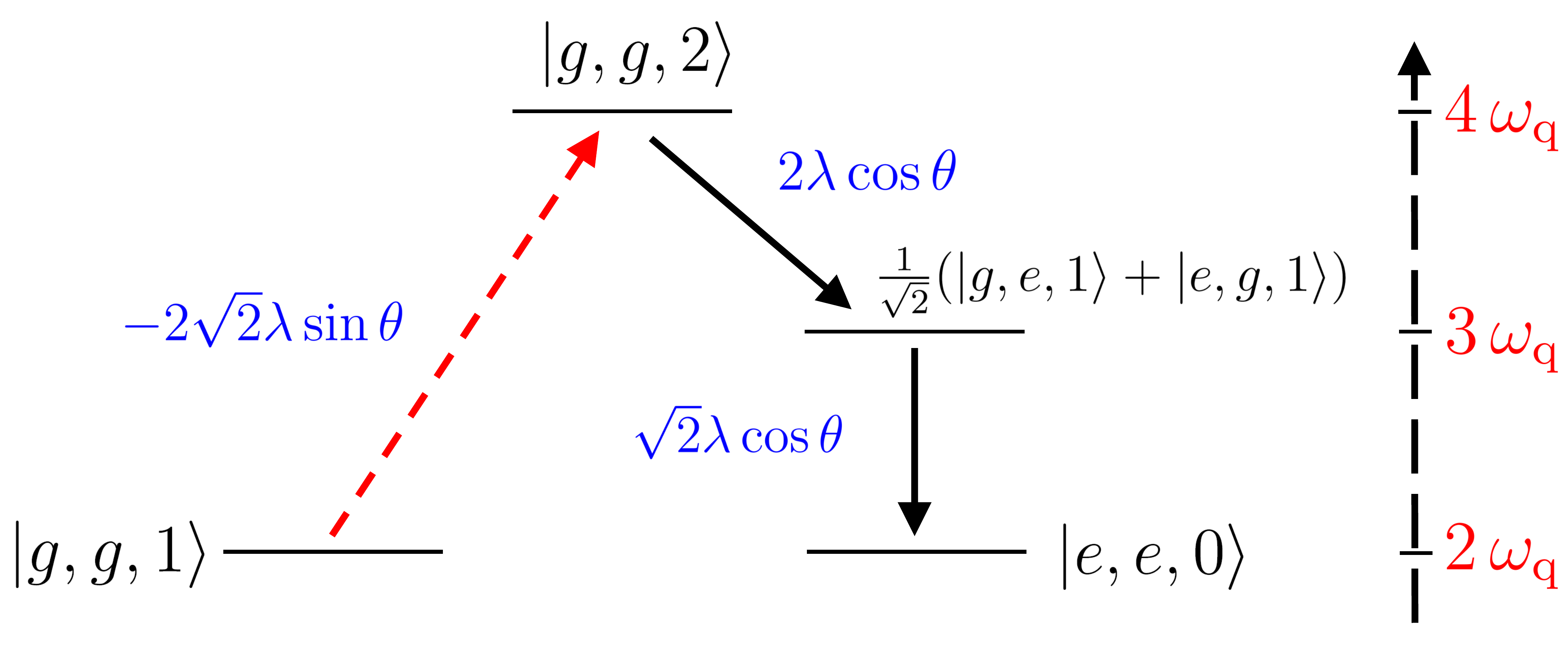}
	\caption{(Color online) Sketch of the process giving the main contribution to the effective coupling between the bare states $| g,g,1 \rangle$ and $|e,e,0 \rangle$, via intermediate virtual transitions. Here, the excitation-number nonconserving processes are represented by arrowed dashed line. The transition matrix elements are also shown.
	 \label{fig:1}}
\end{figure}

We diagonalize numerically the Hamiltonian in Eq.~(\ref{HH}) for the case of two qubits and indicate the resulting energy eigenvalues and eigenstates as $\hbar \omega_{i}$ and  $| i\rangle$  with $i = 0, 1, \dots$, choosing the labelling of the states  such that $\omega_{k} > \omega_{j}$ for $k>j$.  We use a normalized coupling rate $\lambda / \omega_{\rm q} = 0.1$ and an angle $\theta = \pi/6$.   
Figure~2a shows the frequency differences $\omega_{i,0} = \omega_{i} - \omega_{0}$ for the lowest energy states as a function of the resonator frequency.
Starting from the lowest excited states of the spectrum, a large splitting anticrossing around $\omega_{\rm c} / \omega_{\rm q} =1$ can be observed (see arrows in Fig.~2a). It corresponds to the standard vacuum Rabi splitting, which appears also when neglecting the counter-rotating terms. The straight line at $E/  \omega_{\rm q} =1$ corresponds to the dark antisymmetric state $(|g,e,0 \rangle - |e,g,0 \rangle)/\sqrt{2}$. Even larger splitting anticrossings around $\omega_{\rm c} / \omega_{\rm q} =1$ can be observed at higher $E$ values. These correspond to the second and third rung of the JC ladder. We are interested in the region around $\omega_{\rm c} / \omega_{\rm q} =2$, where the levels 3 and 4  display an apparent crossing at $E / \omega_{\rm q} \approx 2$. Actually, what appears as a crossing on this scale, it turns out to be a splitting anticrossing on an enlarged view as in Fig.~2b.
Observing that just outside this avoided-crossing region one level remains flat as a function of $\omega_{\rm c}$ with energy $\omega \approx 2 \omega_{\rm q}$, while the other grows as $\omega_{\rm c}$, this splitting clearly originates from the hybridization of the states $|e,e,0 \rangle$ and $|g,g, 1 \rangle$. The resulting states are well approximated by the states $(|e,e,0 \rangle  \pm |g,g, 1 \rangle)/\sqrt{2}$.
 This splitting is not present in the rotating-wave approximation (RWA), where the coherent coupling between states with a different number of excitations is not allowed, nor does it occur in the absence of symmetry breaking ($\theta =0$). The normalized splitting has a value 2 $\Omega_{\rm eff } / \omega_{\rm q} = 1.97 \times 10^{-3}$, which is in good agreement with  $2 \times 10^{-3} $, obtained within perturbation theory. This observed hybridization opens the way to the observation of weird effects such as the simultaneous excitations of \textit{two} qubits with only \textit{one} cavity photon. Such a coupling between the states $|e,e,0 \rangle$ and $|g,g,1 \rangle$ can be analytically described by the effective interaction Hamiltonian $H_{\rm eff} = -\Omega_{\rm eff} (|e,e,0 \rangle \langle g,g,1| + {\rm H.c.})$.
\begin{figure}[!ht]
	\includegraphics[scale=.40]{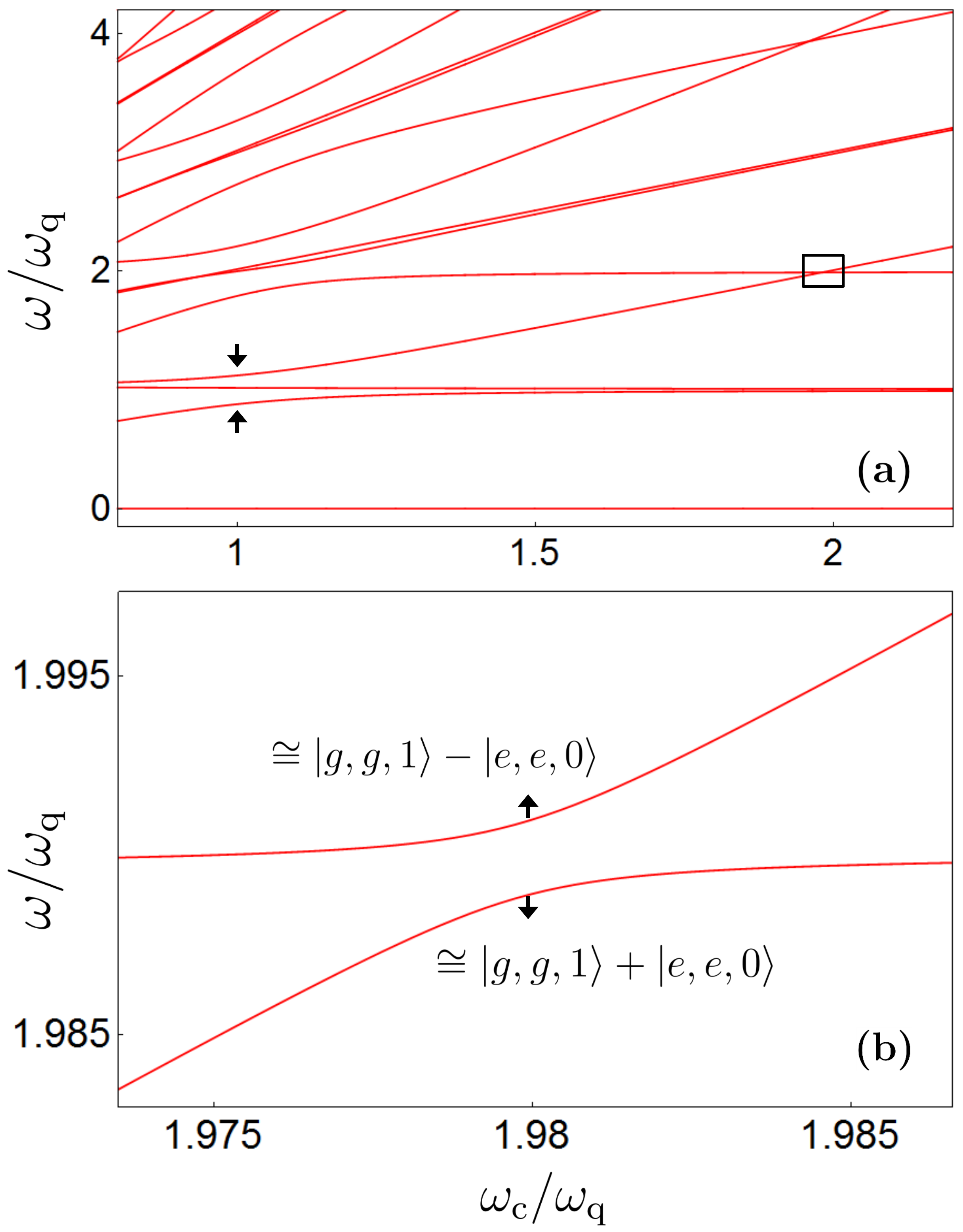}
	\caption{(Color online) (a) Frequency differences $\omega_{i,0} = \omega_{i} - \omega_{0}$ for the lowest energy eigenstates of Hamiltonian~(\ref{HH}) as a function of $\omega_{\rm c}/ \omega_{\rm q}$. Here we consider a normalized coupling rate $\lambda/\omega_{\rm q}=0.1$ between the resonator and each of the qubits. We used $\theta = \pi/6$. The black arrows indicate the ordinary vacuum splitting arising from the coupling between the states $|g,g,1 \rangle$ and $(1 /\sqrt{2}) ( |g,e,0 \rangle + |e,g,0 \rangle)$.
		(b) Enlarged view of the spectral region delimited by a square in panel (a). This shows an avoided-level crossing,  demonstrating  the coupling between the states $|g,g,1 \rangle$ and $|e,e,0 \rangle$ due to the presence of counter-rotating terms in the system Hamiltonian. \label{fig:2PhEnergyLevels}}
\end{figure}

A key theoretical issue of the USC regime is the distinction between bare (unobservable) excitations and physical particles that can be detected \cite{Ridolfo2012,Stassi2013}.
For example, when the counter-rotating terms are taken into account, the mean photon number in the system ground state becomes different from zero: $\langle 0|\hat a^{\dagger} \hat a|0 \rangle\neq 0$.
However, these photons are actually virtual \cite{Ridolfo2012} because they do not correspond to real particles that can be detected in a photon-counting experiment. The same problem holds for the excited states.
According to these analyses, the presence of an $n$-photon contribution in a specific eigenstate of the system does not imply that the system can emit $n$ photons when prepared in this state.

In order to fully understand and characterize this anomalous avoided crossing not present in the RWA, a more quantitative analysis is required.
In the following, we therefore calculate the output signals and correlations which can be measured in a photodetection experiment.
We fix the cavity frequency at the value where the splitting between level 3 and 4 is minimum. Instead of starting from the ideal initial state $(|3 \rangle - |4 \rangle)/\sqrt{2} \approx |g,g,1 \rangle$, more realistically, we consider the system initially 
in its ground state $|0 \rangle \approx |g,g,0 \rangle$ and
 study the direct excitation of the cavity by an electromagnetic Gaussian pulse with central frequency $\omega_{\rm d} = (\omega_{4,0} + \omega_{3,0})/2$.
In this strongly-dispersive regime, the resonator displays very low anharmonicity, so that for a strong system excitation as that induced by a $\pi$-pulse, higher-energy states of the resonator (as the state $| 8 \rangle \simeq |g,g,2 \rangle$) can be resonantly populated.
This problem can be avoided by feeding the system with a single photon input or by probing the system in the weak-excitation regime. However, in order to achieve a deterministic transition $| g,g,1 \rangle \to |e,e,0 \rangle$, a useful route involves introducing a Kerr nonlinearity into the resonator, able to activate a photon blockade.  In circuit QED this can be realized by introducing some additional Josephson junction, or coupling the resonator with weakly-detuned artificial atoms  \cite{Hoffman2011}. This additional nonlinearity can be described by the Hamiltonian term $\hat H_{\rm K} = \mu\,  \hat a^{\dag\, 2}\,  \hat a^2$.
The driving Hamiltonian, describing the system excitation by a coherent electromagnetic pulse is $
\hat H_{\rm d}(t) = {\cal E}(t) \cos (\omega t) \hat X$, 
where ${\cal E}(t) = A \exp{[-(t-t_0)^2/(2 \tau^2)]}/(\tau \sqrt{2 \pi})$. Here, $A$ and $\tau$ are the amplitude  and the standard deviation of the Gaussian pulse, respectively.
$A$ includes the factor $\sqrt{\kappa}$, where $\kappa$ is the loss rate through the cavity port.
The system is thus under the influence of the total Hamiltonian $\hat H = \hat H_0 + \hat H_{\rm K} + \hat H_{\rm d}(t)$.

  The output photon flux emitted by a resonator can be expressed as $\Phi_{\rm out}= \kappa \langle \hat X^{-} \hat X^{+}\rangle$, where $\hat X^{+}=\sum_{j,k>j}X_{jk}|j\rangle \langle k|$ and $\hat X^{-}=(\hat X^{+})^{\dagger}$, with $X_{jk}\equiv \langle j| (\hat a^{\dagger}+\hat a)|k\rangle$, are the positive and negative frequency cavity-photon operators \cite{Garziano2013,Garziano2015}. Neglecting the counter-rotating terms, or in the limit of negligible coupling rates, they coincide with $\hat a$ and $\hat a^\dag$, respectively. 
The signal directly emitted from the qubit is proportional to the qubit mean excitation number $\langle \hat C^- \hat C^+ \rangle$, where $\hat C^\pm$ are the qubit positive and negative frequency operators, defined as $\hat C^{+}=\sum_{j,k>j}C_{jk}|j\rangle \langle k|$ \;and \; $\hat C^{-}=(\hat C^{+})^{\dagger}$, with $C_{jk}\equiv \langle j|(\hat \sigma_{-}+ \hat \sigma_{+})|k\rangle$ \cite{Garziano2013,Garziano2015}.  Neglecting the counter-rotating terms, or in the limit of negligible coupling rates, they coincide with $\hat \sigma_-$ and $\hat \sigma_+$, respectively. In circuit QED systems, this emission can be detected by coupling the qubit to an additional microwave antenna \cite{Hofheinz2009}.

Thanks to the photon-blockade effect, induced by the Kerr interaction $\hat H_{\rm K}$, it is possible to resonantly excite the split states $| 3 \rangle$ and $|4 \rangle$ with a $\pi$-pulse, so that after the pulse arrival the population is completely transferred from the ground state to only these two energy levels. We use a pulse width $\tau = 1/(4\, \omega_{43})$.
\begin{figure}[ht]
	\includegraphics[height= 100 mm]{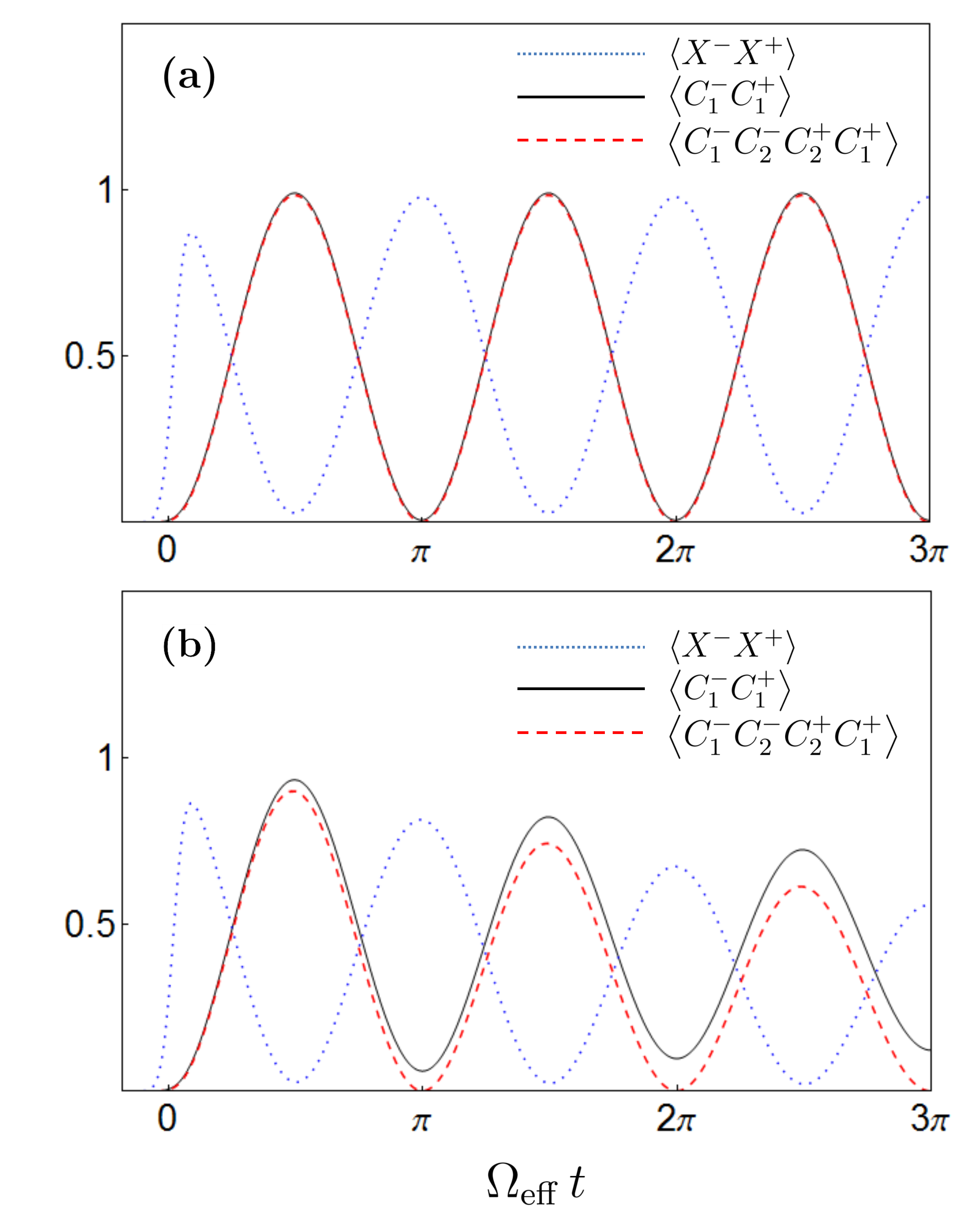}
	\caption{(Color online) (a)  Time evolution of the cavity mean photon number $\langle \hat X^{-} \hat X^{+}\rangle$ (dotted blue curve),  qubit 1 mean excitation number $\langle \hat C_1^{-} \hat C_1^{+}\rangle$ (continuous black curve), and  the zero-delay two-qubit correlation function $G_{\rm q}^{(2)} = \langle \hat C_1^- \hat C_2^-\hat C_2^+ \hat C_1^+\rangle$ (dashed red curve) after the arrival of a $\pi$-like Gaussian pulse initially exciting the resonator. After the arrival of the pulse, the system undergoes vacuum Rabi oscillations showing the reversible joint absorption and re-emission of one photon by two qubits. $\langle \hat C_1^{-} \hat C_1^{+}\rangle$ and $G_{\rm q}^{(2)}(t)$  are almost coincident. This perfect two-qubit correlation is a signature that the two qubits are jointly excited.
(b)	Time evolution of the cavity mean photon number (dotted blue curve),  the qubit  mean excitation number, and the two-qubit correlation  as in (a), but including the effect of cavity damping and atomic decay. The corresponding rates are $\kappa = \gamma = 4 \times 10^{-5} \omega_{\rm q}$.
		} \label{fig:2PhCorrelationFunction}
\end{figure}
 Figure~3a displays the numerically-calculated dynamics of the photon number  $\langle \hat X^{-} \hat X^{+}\rangle$, of the  mean excitation number $\langle \hat C_1^- \hat C_1^+ \rangle$ for qubit 1 (which, of course, coincides with that of qubit 2), and of the two-qubit correlation $G_{\rm q}^{(2)} \equiv \langle \hat C_1^- \hat C_2^-\hat C_2^+ \hat C_1^+\rangle$.
 Vacuum Rabi oscillations showing the reversible excitation exchange between the qubits and the resonator are clearly visible. 
  We observe that, after a half Rabi period, $\Omega_{\rm eff}\, t = \pi/2$, the excitation is fully transferred to the two qubits which reach an excitation probability approaching one. Hence, not only the multiatom absorption of a single photon is possible, but it can  essentially be deterministic. 
  We observe that the single-qubit excitation $\langle \hat C_i^- \hat C_i^+ \rangle$ and $G_{\rm q}^{(2)}$  almost coincide at any time. This almost-perfect two-qubit correlation is a clear signature of the joint excitation: if one qubit gets excited, the probability that also the other one is excited is very close to one.
  In summary, an electromagnetic pulse is able, thanks to the photon blockade effect, to generate a \textit{single} cavity-photon, which then gets jointly absorbed by a \textit{couple} of qubits. 
  The resonant coupling can be stopped at this time, {e.g.}, by changing the resonance frequency of the qubits. If not, the reverse process starts, where two qubits jointly emit a  \textit{single} photon:  $|e,e,0 \rangle \to |g,g,1 \rangle$. 
  We observe that  $\langle \hat X^{-} \hat X^{+}\rangle$ is not exactly zero at the photon minima. This occurs because the two-qubit excited state, owing to the same processes inducing its coupling with the one-photon state, acquires a dipole transition matrix element, so that this state is able to emit photons. We find that (not shown here) this effects increase when increasing the atom-field coupling strength $\lambda$.
   In order to exclude that this joint qubit excitation does not occur via more conventional paths, involving the creation of photon pairs and/or a 1-qubit-1-photon excitation, we have also calculated the photonic second-order correlation function  $G_{\rm c}^{(2)} \equiv \langle (\hat X^-)^2 (\hat X^+)^2\rangle$
  and the qubit-cavity correlation  $G_{\rm qc}^{(2)} \equiv \langle \hat C_i^- \hat X^-\hat X^+ \hat C_i^+\rangle$. We find that their value is more than two orders of magnitude lower than that of the two-qubit correlation  $G_{\rm q}^{(2)}$.
  
  Figure~3a has been obtained without including loss effects.
The influence of cavity field damping and atomic decay on the  process can be studied by the master equation approach. We consider the system interacting with zero-temperature baths.
By using the Born-Markov approximation without the post-trace RWA \cite{Law2015},  the resulting  master equation  for the reduced density matrix of the system is
\begin{equation}\label{ME}
\dot{\hat\rho}= i [\hat \rho(t),\hat H] + \kappa {\cal D}\, [\hat X^+]{\hat\rho}
+ \gamma \sum_i  {\cal D}[\hat C_i^+]\, {\hat\rho}\, ,
\end{equation}
where the superoperator ${\cal D}$ is defined as 
$\mathcal{D}[\hat O]\hat\rho=\frac{1}{2}(2 \hat O\, \hat\rho\, \hat O^{\dagger}-\hat \rho\, \hat O^{\dagger}\, \hat O-\hat O^{\dagger}\, \hat O\,  \hat \rho)$.
We use $\kappa = \gamma = 3 \times 10^{-5}\,  \omega_{\rm q}$. Figure~3b shows how the cavity losses and the atomic decay affects the system dynamics. As expected, the 
vacuum Rabi oscillations undergo damping and, as expected, the two-qubit correlation is more fragile to losses.
Finally, we have also considered the case when the two qubits display different coupling rates with the resonator field. We used $\lambda_1 /\omega_{\rm q}= 8 \times 10^{-2}$ and 
$\lambda_2 /\omega_{\rm q}= 1.2 \times 10^{-1}$. We found that also in this case 
$\langle \hat C_1^- \hat C_1^+ \rangle = \langle \hat C_2^- \hat C_2^+ \rangle \simeq  G_{\rm q}^{(2)}$. This result \textit{further confirms the simultaneous and joint nature of this multiatom process}.

The processes described here can be observed by placing two superconducting artificial atoms at opposite ends of a superconducting transmission line resonator \cite{Majer2007}.
These multi-atom excitation and emission processes can find useful applications for the development of novel quantum technologies. Conditional quantum-state transfer is a first possible application: the quantum information stored in one of the two qubits can be transferred to the resonator conditioned by the state of the second qubit. We also observe that the quantum Rabi oscillations displayed in Fig.~3 imply that  a hybrid entangled GHZ state, $(|g,g,1 \rangle + |e,e,0 \rangle)/ \sqrt{2}$, can be obtained by an elementary quantum Rabi process
after a time $t = \pi/(4 \Omega_{\rm eff})$. This state can be stored by just changing the transition frequency of one of the two qubits.
Besides possible applications, the puzzling results presented here, showing that \textit{one} photon can divide its energy into \textit{two} spatially-separated atoms, and that vacuum fluctuations \cite{Norirmp} can induce separate atoms to  behave as a single quantum entity (as testified by the one-photon transition matrix element acquired by the transition $|g,g \rangle \to |e,e \rangle$), provide new insights into the quantum aspects of the interaction between light and matter.

We hope that this proposal for the
simultaneous excitation of two or three atoms with a single photon, might be effective in producing the
simultaneous excitation of two or three referees with a single manuscript.

This work is partially supported by the RIKEN iTHES Project, the MURI Center for Dynamic Magneto-Optics via the AFOSR award number FA9550-14-1-0040,
the IMPACT program of JST, a Grant-in-Aid for Scientific Research (A), and  from the MPNS
COST Action MP1403 Nanoscale Quantum Optics.
\newline
\newline
\noindent
\begin{center}
{\Large \bf Supplemental Material:}

\vspace{0.7 cm}
\noindent {\bf DERIVATION OF THE EFFECTIVE HAMILTONIAN}
\end{center}
According to standard time-dependent perturbation theory, for a constant perturbation switched-on at $t=0$, the resulting transition rate can be expressed as
\begin{equation}
	W_{i \to f} = \frac{2 \pi}{\hbar} |V^{\rm eff}_{f i}|^2\,  \delta(E_f - E_i)\, ,
\end{equation}
where $i$ and $f$ label the initial and final states with corresponding energies $E_i$ and $E_f$,  and $V^{\rm eff}_{f\, i}$ describes the effective coupling strength connecting the initial and final states. In the framework of first-order perturbation theory, this effective coupling strength coincides with the matrix element of the generic  perturbing interaction $\hat V$: $V^{\rm eff}_{f i} = V_{f i}= \langle f | \hat V | i \rangle$.
If $i$ and $f$ are coupled only via third-order perturbation theory, the resulting effective coupling strength is,
\begin{equation}\label{eff}
	V^{\rm eff}_{f i} =  \sum_{m,n} \frac{V_{fn} V_{nm} V_{mi}}{(E_i - E_m)(E_i - E_n)}\, .
\end{equation}

In the case when the states $|n \rangle$ and $|m \rangle$  are virtual intermediate states that do not conserve energy, the only effect of the perturbation is to couple, via these virtual intermediate states, the initial and final states. The same coupling can be described by the effective Hamiltonian
\begin{equation}
	H_{\rm eff} = V^{\rm eff}_{f i}\, | f \rangle \langle i | + {\rm H. c.}\, ,
\end{equation}
with $V^{\rm eff}_{f i}$ provided by Eq.~(2).

We observe that, applying first-order perturbation theory by using this effective Hamiltonian, we obtain the same result of standard  third-order perturbation theory with the real perturbation $\hat V$. Hence, Eq.~(\ref{eff}) describes the effective coupling strength between the energy-degenerate states $| g,g,1 \rangle$ and
$|e,e,0\rangle$. We consider the case $\omega_{\rm c} \approx 2 \omega_{\rm q}$ and a perturbation of the form:
\begin{equation}
	\hat V =\lambda  \hat X
	\sum_i
	(\cos \theta\,  \hat \sigma^{(i)}_x + \sin \theta\, \hat \sigma^{(i)}_z )\, .
	\label{Hpert}
\end{equation}
\begin{figure}[!ht]
	\centering
	\includegraphics[height=19.8cm,width=8.5cm]{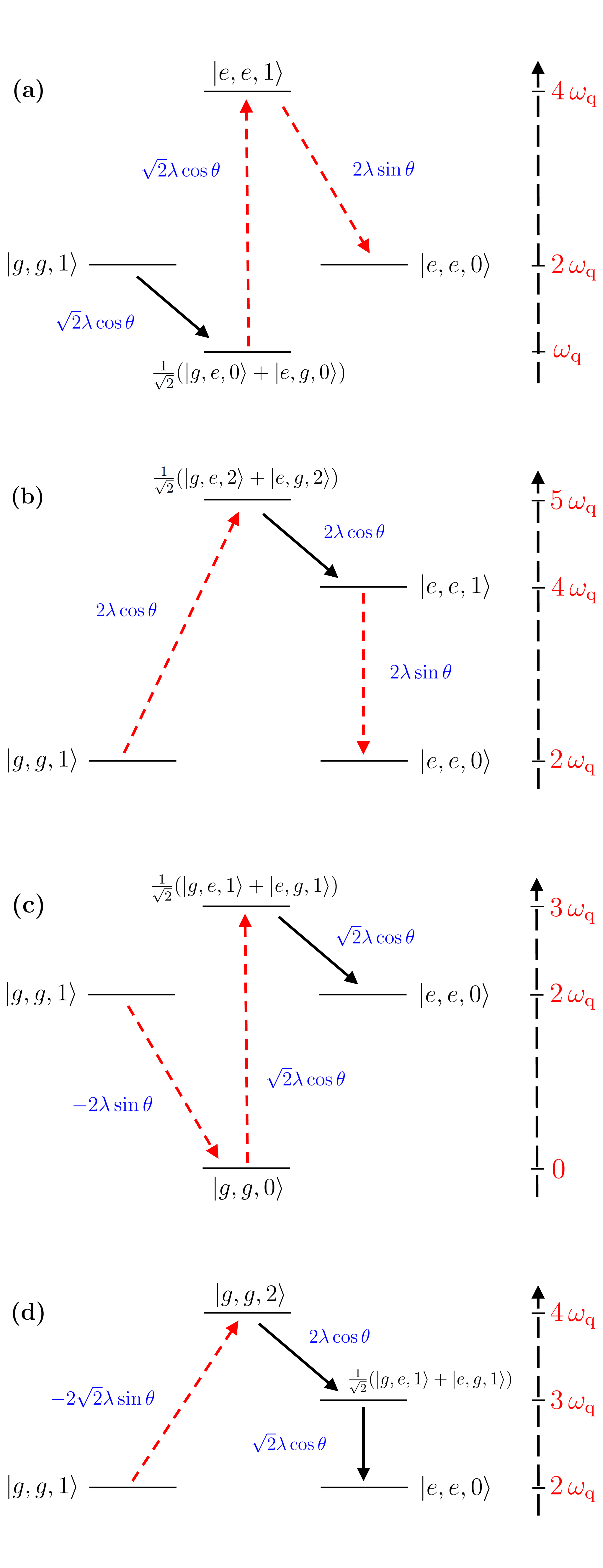}
	\caption{(Color online) Coupling between the bare states $| g,g,1 \rangle$ and $|e,e,0 \rangle$ via intermediate virtual transitions. Here, the excitation-number nonconserving processes are represented by the arrowed red dashed lines. The transition matrix elements are also shown (in blue).
		\label{fig:1}}
\end{figure}

After carefully inspecting all the possible intermediate states, we find that only the four paths shown in Fig.~S1 can connect the states $| g,g,1 \rangle$ and
$|e,e,0\rangle$.
Applying Eq.~(\ref{eff}), we obtain,
\begin{equation}\label{eff1}
	\Omega_{\rm eff} \equiv -  V^{\rm eff}_{f i} = \frac{8}{3}\sin \theta \cos^2{\theta}  \left(\frac{\lambda}{\omega_{\rm q}}\right)^3 \, .
\end{equation}
\begin{figure}[ht!]
	\includegraphics[height=70 mm]{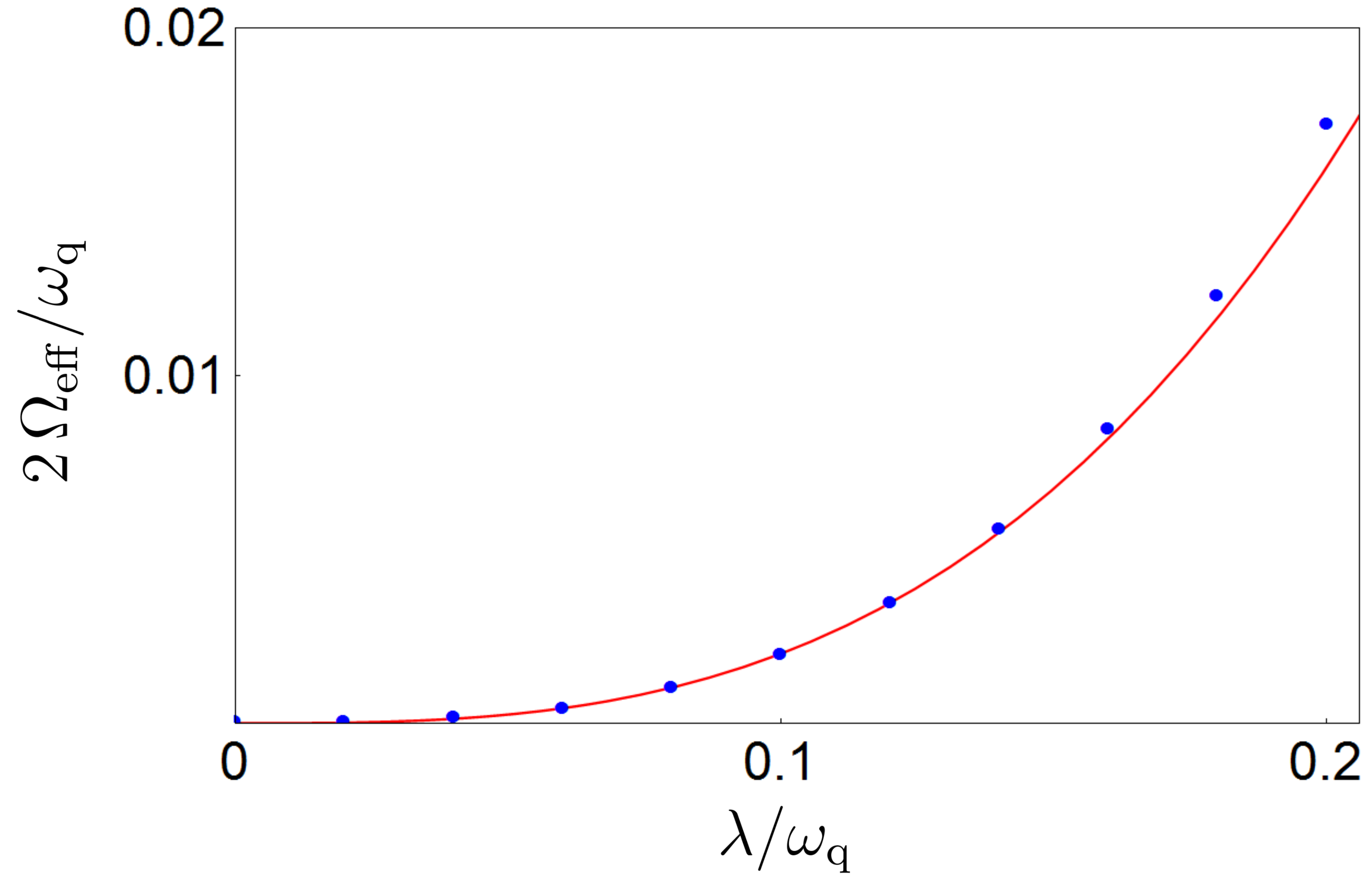}
	\caption{(Color online) Comparison between the numerically-calculated normalized Rabi splitting (points)  (corresponding to twice the effective coupling between one cavity photon and two independent atoms) and the corresponding calculation using third-order perturbation theory (continuous red curve).
	}\label{Fig2}
\end{figure}
Figure~S2 displays the comparison of the magnitudes of the effective Rabi splitting $2  \Omega_{\rm eff}/ \omega_{\rm q} $  between the states  $| g,g,1 \rangle$ and
$|e,e,0\rangle$ obtained analytically [Eq.~(5)] via third-order perturbation theory
and by the numerical diagonalization of the Hamiltonian in Eq.~(1) (in the main text), as a function of  the normalized interaction strength $ \lambda/\omega_{\rm q} $.
The agreement is very good, also  for coupling strengths $\lambda$ beyond $10 \%$ of the qubit transition frequency $\omega_{\rm q}$. This result confirms the $(\lambda / \omega_{\rm q})^3$ proportionality of the effective (one-photon)-(two-atoms) coupling predicted by the above analysis.

\newpage

\bibliography{Riken}

\begin{thebibliography}{43}%
\makeatletter
\providecommand \@ifxundefined [1]{%
 \@ifx{#1\undefined}
}%
\providecommand \@ifnum [1]{%
 \ifnum #1\expandafter \@firstoftwo
 \else \expandafter \@secondoftwo
 \fi
}%
\providecommand \@ifx [1]{%
 \ifx #1\expandafter \@firstoftwo
 \else \expandafter \@secondoftwo
 \fi
}%
\providecommand \natexlab [1]{#1}%
\providecommand \enquote  [1]{``#1''}%
\providecommand \bibnamefont  [1]{#1}%
\providecommand \bibfnamefont [1]{#1}%
\providecommand \citenamefont [1]{#1}%
\providecommand \href@noop [0]{\@secondoftwo}%
\providecommand \href [0]{\begingroup \@sanitize@url \@href}%
\providecommand \@href[1]{\@@startlink{#1}\@@href}%
\providecommand \@@href[1]{\endgroup#1\@@endlink}%
\providecommand \@sanitize@url [0]{\catcode `\\12\catcode `\$12\catcode
  `\&12\catcode `\#12\catcode `\^12\catcode `\_12\catcode `\%12\relax}%
\providecommand \@@startlink[1]{}%
\providecommand \@@endlink[0]{}%
\providecommand \url  [0]{\begingroup\@sanitize@url \@url }%
\providecommand \@url [1]{\endgroup\@href {#1}{\urlprefix }}%
\providecommand \urlprefix  [0]{URL }%
\providecommand \Eprint [0]{\href }%
\providecommand \doibase [0]{http://dx.doi.org/}%
\providecommand \selectlanguage [0]{\@gobble}%
\providecommand \bibinfo  [0]{\@secondoftwo}%
\providecommand \bibfield  [0]{\@secondoftwo}%
\providecommand \translation [1]{[#1]}%
\providecommand \BibitemOpen [0]{}%
\providecommand \bibitemStop [0]{}%
\providecommand \bibitemNoStop [0]{.\EOS\space}%
\providecommand \EOS [0]{\spacefactor3000\relax}%
\providecommand \BibitemShut  [1]{\csname bibitem#1\endcsname}%
\let\auto@bib@innerbib\@empty
\bibitem [{\citenamefont {G{\"o}ppert-Mayer}(1931)}]{Goeppert-Mayer1931}%
  \BibitemOpen
  \bibfield  {author} {\bibinfo {author} {\bibfnamefont {M.}~\bibnamefont
  {G{\"o}ppert-Mayer}},\ }\bibfield  {title} {\enquote {\bibinfo {title}
  {{\"U}ber elementarakte mit zwei quantenspr{\"u}ngen},}\ }\href@noop {}
  {\bibfield  {journal} {\bibinfo  {journal} {Annalen der Physik}\ }\textbf
  {\bibinfo {volume} {401}},\ \bibinfo {pages} {273--294} (\bibinfo {year}
  {1931})}\BibitemShut {NoStop}%
\bibitem [{\citenamefont {Denk}\ \emph {et~al.}(1990)\citenamefont {Denk},
  \citenamefont {Strickler},\ and\ \citenamefont {Webb}}]{Denk1990}%
  \BibitemOpen
  \bibfield  {author} {\bibinfo {author} {\bibfnamefont {W.}~\bibnamefont
  {Denk}}, \bibinfo {author} {\bibfnamefont {J.~H.}\ \bibnamefont {Strickler}},
  \ and\ \bibinfo {author} {\bibfnamefont {W.~W}\ \bibnamefont {Webb}},\
  }\bibfield  {title} {\enquote {\bibinfo {title} {Two-photon laser scanning
  fluorescence microscopy},}\ }\href@noop {} {\bibfield  {journal} {\bibinfo
  {journal} {Science}\ }\textbf {\bibinfo {volume} {248}},\ \bibinfo {pages}
  {73--76} (\bibinfo {year} {1990})}\BibitemShut {NoStop}%
\bibitem [{\citenamefont {So}\ \emph {et~al.}(2000)\citenamefont {So},
  \citenamefont {Dong}, \citenamefont {Masters},\ and\ \citenamefont
  {Berland}}]{So2000}%
  \BibitemOpen
  \bibfield  {author} {\bibinfo {author} {\bibfnamefont {P.~T.~C.}\
  \bibnamefont {So}}, \bibinfo {author} {\bibfnamefont {C.~Y.}\ \bibnamefont
  {Dong}}, \bibinfo {author} {\bibfnamefont {B.~R.}\ \bibnamefont {Masters}}, \
  and\ \bibinfo {author} {\bibfnamefont {K.~M.}\ \bibnamefont {Berland}},\
  }\bibfield  {title} {\enquote {\bibinfo {title} {Two-photon excitation
  fluorescence microscopy},}\ }\href@noop {} {\bibfield  {journal} {\bibinfo
  {journal} {Annual review of biomedical engineering}\ }\textbf {\bibinfo
  {volume} {2}},\ \bibinfo {pages} {399--429} (\bibinfo {year}
  {2000})}\BibitemShut {NoStop}%
\bibitem [{\citenamefont {Haroche}\ and\ \citenamefont
  {Raimond}(2006)}]{Haroche2006}%
  \BibitemOpen
  \bibfield  {author} {\bibinfo {author} {\bibfnamefont {S.}~\bibnamefont
  {Haroche}}\ and\ \bibinfo {author} {\bibfnamefont {J.~M.}\ \bibnamefont
  {Raimond}},\ }\href@noop {} {\emph {\bibinfo {title} {Exploring the quantum:
  Atoms, Cavities and Photons}}}\ (\bibinfo  {publisher} {Oxford Univ. Press},\
  \bibinfo {year} {2006})\BibitemShut {NoStop}%
\bibitem [{\citenamefont {Haroche}(2013)}]{Haroche2013}%
  \BibitemOpen
  \bibfield  {author} {\bibinfo {author} {\bibfnamefont {S.}~\bibnamefont
  {Haroche}},\ }\bibfield  {title} {\enquote {\bibinfo {title} {Nobel lecture:
  Controlling photons in a box and exploring the quantum to classical
  boundary},}\ }\href@noop {} {\bibfield  {journal} {\bibinfo  {journal} {Rev.
  Mod. Phys.}\ }\textbf {\bibinfo {volume} {85}},\ \bibinfo {pages} {1083}
  (\bibinfo {year} {2013})}\BibitemShut {NoStop}%
\bibitem [{\citenamefont {Haroche}\ \emph {et~al.}(2013)\citenamefont
  {Haroche}, \citenamefont {Brune},\ and\ \citenamefont
  {Raimond}}]{Haroche2013a}%
  \BibitemOpen
  \bibfield  {author} {\bibinfo {author} {\bibfnamefont {S.}~\bibnamefont
  {Haroche}}, \bibinfo {author} {\bibfnamefont {M.}~\bibnamefont {Brune}}, \
  and\ \bibinfo {author} {\bibfnamefont {J.~M.}\ \bibnamefont {Raimond}},\
  }\bibfield  {title} {\enquote {\bibinfo {title} {Atomic clocks for
  controlling light fields},}\ }\href@noop {} {\bibfield  {journal} {\bibinfo
  {journal} {Physics Today}\ }\textbf {\bibinfo {volume} {66}},\ \bibinfo
  {pages} {27} (\bibinfo {year} {2013})}\BibitemShut {NoStop}%
\bibitem [{\citenamefont {Deleglise}\ \emph {et~al.}(2008)\citenamefont
  {Deleglise}, \citenamefont {Dotsenko}, \citenamefont {Sayrin}, \citenamefont
  {Bernu}, \citenamefont {Brune}, \citenamefont {Raimond},\ and\ \citenamefont
  {Haroche}}]{Deleglise2008}%
  \BibitemOpen
  \bibfield  {author} {\bibinfo {author} {\bibfnamefont {S.}~\bibnamefont
  {Deleglise}}, \bibinfo {author} {\bibfnamefont {I.}~\bibnamefont {Dotsenko}},
  \bibinfo {author} {\bibfnamefont {C.}~\bibnamefont {Sayrin}}, \bibinfo
  {author} {\bibfnamefont {J.}~\bibnamefont {Bernu}}, \bibinfo {author}
  {\bibfnamefont {M.}~\bibnamefont {Brune}}, \bibinfo {author} {\bibfnamefont
  {J.~M.}\ \bibnamefont {Raimond}}, \ and\ \bibinfo {author} {\bibfnamefont
  {S.}~\bibnamefont {Haroche}},\ }\bibfield  {title} {\enquote {\bibinfo
  {title} {Reconstruction of non-classical cavity field states with snapshots
  of their decoherence},}\ }\href@noop {} {\bibfield  {journal} {\bibinfo
  {journal} {Nature}\ }\textbf {\bibinfo {volume} {455}},\ \bibinfo {pages}
  {510--514} (\bibinfo {year} {2008})}\BibitemShut {NoStop}%
\bibitem [{\citenamefont {Hofheinz}\ \emph {et~al.}(2009)\citenamefont
  {Hofheinz}, \citenamefont {Wang}, \citenamefont {Ansmann}, \citenamefont
  {Bialczak}, \citenamefont {Lucero}, \citenamefont {Neeley}, \citenamefont
  {O'Connell}, \citenamefont {Sank}, \citenamefont {Wenner}, \citenamefont
  {Martinis},\ and\ \citenamefont {Cleland}}]{Hofheinz2009}%
  \BibitemOpen
  \bibfield  {author} {\bibinfo {author} {\bibfnamefont {M.}~\bibnamefont
  {Hofheinz}}, \bibinfo {author} {\bibfnamefont {H.}~\bibnamefont {Wang}},
  \bibinfo {author} {\bibfnamefont {M.}~\bibnamefont {Ansmann}}, \bibinfo
  {author} {\bibfnamefont {R.~C.}\ \bibnamefont {Bialczak}}, \bibinfo {author}
  {\bibfnamefont {E.}~\bibnamefont {Lucero}}, \bibinfo {author} {\bibfnamefont
  {M.}~\bibnamefont {Neeley}}, \bibinfo {author} {\bibfnamefont {A.~D.}\
  \bibnamefont {O'Connell}}, \bibinfo {author} {\bibfnamefont {D.}~\bibnamefont
  {Sank}}, \bibinfo {author} {\bibfnamefont {J.}~\bibnamefont {Wenner}},
  \bibinfo {author} {\bibfnamefont {J.~M.}\ \bibnamefont {Martinis}}, \ and\
  \bibinfo {author} {\bibfnamefont {A.~N.}\ \bibnamefont {Cleland}},\
  }\bibfield  {title} {\enquote {\bibinfo {title} {Synthesizing arbitrary
  quantum states in a superconducting resonator},}\ }\href@noop {} {\bibfield
  {journal} {\bibinfo  {journal} {Nature}\ }\textbf {\bibinfo {volume} {459}},\
  \bibinfo {pages} {546--549} (\bibinfo {year} {2009})}\BibitemShut {NoStop}%
\bibitem [{\citenamefont {Vlastakis}\ \emph {et~al.}(2013)\citenamefont
  {Vlastakis}, \citenamefont {Kirchmair}, \citenamefont {Leghtas},
  \citenamefont {Nigg}, \citenamefont {Frunzio}, \citenamefont {Girvin},
  \citenamefont {Mirrahimi}, \citenamefont {Devoret},\ and\ \citenamefont
  {Schoelkopf}}]{Vlastakis2013}%
  \BibitemOpen
  \bibfield  {author} {\bibinfo {author} {\bibfnamefont {B.}~\bibnamefont
  {Vlastakis}}, \bibinfo {author} {\bibfnamefont {G.}~\bibnamefont
  {Kirchmair}}, \bibinfo {author} {\bibfnamefont {Z.}~\bibnamefont {Leghtas}},
  \bibinfo {author} {\bibfnamefont {S.~E.}\ \bibnamefont {Nigg}}, \bibinfo
  {author} {\bibfnamefont {L.}~\bibnamefont {Frunzio}}, \bibinfo {author}
  {\bibfnamefont {S.~M.}\ \bibnamefont {Girvin}}, \bibinfo {author}
  {\bibfnamefont {M.}~\bibnamefont {Mirrahimi}}, \bibinfo {author}
  {\bibfnamefont {M.~H.}\ \bibnamefont {Devoret}}, \ and\ \bibinfo {author}
  {\bibfnamefont {R.~J.}\ \bibnamefont {Schoelkopf}},\ }\bibfield  {title}
  {\enquote {\bibinfo {title} {Deterministically encoding quantum information
  using 100-photon {S}chr{\"o}dinger cat states},}\ }\href@noop {} {\bibfield
  {journal} {\bibinfo  {journal} {Science}\ }\textbf {\bibinfo {volume}
  {342}},\ \bibinfo {pages} {607--610} (\bibinfo {year} {2013})}\BibitemShut
  {NoStop}%
\bibitem [{\citenamefont {Rauschenbeutel}\ \emph {et~al.}(1999)\citenamefont
  {Rauschenbeutel}, \citenamefont {Nogues}, \citenamefont {Osnaghi},
  \citenamefont {Bertet}, \citenamefont {Brune}, \citenamefont {Raimond},\ and\
  \citenamefont {Haroche}}]{Rauschenbeutel1999}%
  \BibitemOpen
  \bibfield  {author} {\bibinfo {author} {\bibfnamefont {A.}~\bibnamefont
  {Rauschenbeutel}}, \bibinfo {author} {\bibfnamefont {G.}~\bibnamefont
  {Nogues}}, \bibinfo {author} {\bibfnamefont {S.}~\bibnamefont {Osnaghi}},
  \bibinfo {author} {\bibfnamefont {P.}~\bibnamefont {Bertet}}, \bibinfo
  {author} {\bibfnamefont {M.}~\bibnamefont {Brune}}, \bibinfo {author}
  {\bibfnamefont {J.~M.}\ \bibnamefont {Raimond}}, \ and\ \bibinfo {author}
  {\bibfnamefont {S.}~\bibnamefont {Haroche}},\ }\bibfield  {title} {\enquote
  {\bibinfo {title} {Coherent operation of a tunable quantum phase gate in
  cavity {Q}{E}{D}},}\ }\href@noop {} {\bibfield  {journal} {\bibinfo
  {journal} {Phys. Rev. Lett.}\ }\textbf {\bibinfo {volume} {83}},\ \bibinfo
  {pages} {5166} (\bibinfo {year} {1999})}\BibitemShut {NoStop}%
\bibitem [{\citenamefont {Zheng}\ and\ \citenamefont {Guo}(2000)}]{Zheng2000}%
  \BibitemOpen
  \bibfield  {author} {\bibinfo {author} {\bibfnamefont {S.-B.}\ \bibnamefont
  {Zheng}}\ and\ \bibinfo {author} {\bibfnamefont {G.-C.}\ \bibnamefont
  {Guo}},\ }\bibfield  {title} {\enquote {\bibinfo {title} {Efficient scheme
  for two-atom entanglement and quantum information processing in cavity
  {Q}{E}{D}},}\ }\href@noop {} {\bibfield  {journal} {\bibinfo  {journal}
  {Phys. Rev. Lett.}\ }\textbf {\bibinfo {volume} {85}},\ \bibinfo {pages}
  {2392} (\bibinfo {year} {2000})}\BibitemShut {NoStop}%
\bibitem [{\citenamefont {You}\ and\ \citenamefont {Nori}(2005)}]{You2006}%
  \BibitemOpen
  \bibfield  {author} {\bibinfo {author} {\bibfnamefont {J.~Q.}\ \bibnamefont
  {You}}\ and\ \bibinfo {author} {\bibfnamefont {F.}~\bibnamefont {Nori}},\
  }\bibfield  {title} {\enquote {\bibinfo {title} {Superconducting circuits and
  quantum information},}\ }\href@noop {} {\bibfield  {journal} {\bibinfo
  {journal} {Phys. Today}\ }\textbf {\bibinfo {volume} {58 (II)}},\ \bibinfo
  {pages} {42} (\bibinfo {year} {2005})}\BibitemShut {NoStop}%
\bibitem [{\citenamefont {Felinto}\ \emph {et~al.}(2006)\citenamefont
  {Felinto}, \citenamefont {Chou}, \citenamefont {Laurat}, \citenamefont
  {Schomburg}, \citenamefont {De~Riedmatten},\ and\ \citenamefont
  {Kimble}}]{Felinto2006}%
  \BibitemOpen
  \bibfield  {author} {\bibinfo {author} {\bibfnamefont {D.}~\bibnamefont
  {Felinto}}, \bibinfo {author} {\bibfnamefont {C.-W.}\ \bibnamefont {Chou}},
  \bibinfo {author} {\bibfnamefont {J.}~\bibnamefont {Laurat}}, \bibinfo
  {author} {\bibfnamefont {E.~W.}\ \bibnamefont {Schomburg}}, \bibinfo {author}
  {\bibfnamefont {H.}~\bibnamefont {De~Riedmatten}}, \ and\ \bibinfo {author}
  {\bibfnamefont {H.~J.}\ \bibnamefont {Kimble}},\ }\bibfield  {title}
  {\enquote {\bibinfo {title} {Conditional control of the quantum states of
  remote atomic memories for quantum networking},}\ }\href@noop {} {\bibfield
  {journal} {\bibinfo  {journal} {Nature Physics}\ }\textbf {\bibinfo {volume}
  {2}},\ \bibinfo {pages} {844--848} (\bibinfo {year} {2006})}\BibitemShut
  {NoStop}%
\bibitem [{\citenamefont {Kimble}(2008)}]{Kimble2008}%
  \BibitemOpen
  \bibfield  {author} {\bibinfo {author} {\bibfnamefont {H.~J.}\ \bibnamefont
  {Kimble}},\ }\bibfield  {title} {\enquote {\bibinfo {title} {The quantum
  internet},}\ }\href@noop {} {\bibfield  {journal} {\bibinfo  {journal}
  {Nature}\ }\textbf {\bibinfo {volume} {453}},\ \bibinfo {pages} {1023--1030}
  (\bibinfo {year} {2008})}\BibitemShut {NoStop}%
\bibitem [{\citenamefont {Ritter}\ \emph {et~al.}(2012)\citenamefont {Ritter},
  \citenamefont {N{\"o}lleke}, \citenamefont {Hahn}, \citenamefont {Reiserer},
  \citenamefont {Neuzner}, \citenamefont {Uphoff}, \citenamefont {M{\"u}cke},
  \citenamefont {Figueroa}, \citenamefont {Bochmann},\ and\ \citenamefont
  {Rempe}}]{Ritter2012}%
  \BibitemOpen
  \bibfield  {author} {\bibinfo {author} {\bibfnamefont {S.}~\bibnamefont
  {Ritter}}, \bibinfo {author} {\bibfnamefont {C.}~\bibnamefont {N{\"o}lleke}},
  \bibinfo {author} {\bibfnamefont {C.}~\bibnamefont {Hahn}}, \bibinfo {author}
  {\bibfnamefont {A.}~\bibnamefont {Reiserer}}, \bibinfo {author}
  {\bibfnamefont {A.}~\bibnamefont {Neuzner}}, \bibinfo {author} {\bibfnamefont
  {M.}~\bibnamefont {Uphoff}}, \bibinfo {author} {\bibfnamefont
  {M.}~\bibnamefont {M{\"u}cke}}, \bibinfo {author} {\bibfnamefont
  {E.}~\bibnamefont {Figueroa}}, \bibinfo {author} {\bibfnamefont
  {J.}~\bibnamefont {Bochmann}}, \ and\ \bibinfo {author} {\bibfnamefont
  {G.}~\bibnamefont {Rempe}},\ }\bibfield  {title} {\enquote {\bibinfo {title}
  {An elementary quantum network of single atoms in optical cavities},}\
  }\href@noop {} {\bibfield  {journal} {\bibinfo  {journal} {Nature}\ }\textbf
  {\bibinfo {volume} {484}},\ \bibinfo {pages} {195--200} (\bibinfo {year}
  {2012})}\BibitemShut {NoStop}%
\bibitem [{\citenamefont {You}\ and\ \citenamefont {Nori}(2003)}]{You2003}%
  \BibitemOpen
  \bibfield  {author} {\bibinfo {author} {\bibfnamefont {J.~Q.}\ \bibnamefont
  {You}}\ and\ \bibinfo {author} {\bibfnamefont {F.}~\bibnamefont {Nori}},\
  }\bibfield  {title} {\enquote {\bibinfo {title} {Quantum information
  processing with superconducting qubits in a microwave field},}\ }\href@noop
  {} {\bibfield  {journal} {\bibinfo  {journal} {Phys. Rev. B}\ }\textbf
  {\bibinfo {volume} {68}},\ \bibinfo {pages} {064509} (\bibinfo {year}
  {2003})}\BibitemShut {NoStop}%
\bibitem [{\citenamefont {Blais}\ \emph {et~al.}(2004)\citenamefont {Blais},
  \citenamefont {Huang}, \citenamefont {Wallraff}, \citenamefont {Girvin},\
  and\ \citenamefont {Schoelkopf}}]{Blais2004}%
  \BibitemOpen
  \bibfield  {author} {\bibinfo {author} {\bibfnamefont {A.}~\bibnamefont
  {Blais}}, \bibinfo {author} {\bibfnamefont {R.~S.}\ \bibnamefont {Huang}},
  \bibinfo {author} {\bibfnamefont {A.}~\bibnamefont {Wallraff}}, \bibinfo
  {author} {\bibfnamefont {S.~M.}\ \bibnamefont {Girvin}}, \ and\ \bibinfo
  {author} {\bibfnamefont {R.~J.}\ \bibnamefont {Schoelkopf}},\ }\bibfield
  {title} {\enquote {\bibinfo {title} {Cavity quantum electrodynamics for
  superconducting electrical circuits: An architecture for quantum
  computation},}\ }\href@noop {} {\bibfield  {journal} {\bibinfo  {journal}
  {Phys. Rev. A}\ }\textbf {\bibinfo {volume} {69}},\ \bibinfo {pages} {062320}
  (\bibinfo {year} {2004})}\BibitemShut {NoStop}%
\bibitem [{\citenamefont {Wallraff}\ \emph {et~al.}(2004)\citenamefont
  {Wallraff}, \citenamefont {Schuster}, \citenamefont {Blais}, \citenamefont
  {Frunzio}, \citenamefont {Huang}, \citenamefont {Majer}, \citenamefont
  {Kumar}, \citenamefont {Girvin},\ and\ \citenamefont
  {Schoelkopf}}]{Wallraff2004}%
  \BibitemOpen
  \bibfield  {author} {\bibinfo {author} {\bibfnamefont {A.}~\bibnamefont
  {Wallraff}}, \bibinfo {author} {\bibfnamefont {D.~I.}\ \bibnamefont
  {Schuster}}, \bibinfo {author} {\bibfnamefont {A.}~\bibnamefont {Blais}},
  \bibinfo {author} {\bibfnamefont {L.}~\bibnamefont {Frunzio}}, \bibinfo
  {author} {\bibfnamefont {R.~S.}\ \bibnamefont {Huang}}, \bibinfo {author}
  {\bibfnamefont {J.}~\bibnamefont {Majer}}, \bibinfo {author} {\bibfnamefont
  {S.}~\bibnamefont {Kumar}}, \bibinfo {author} {\bibfnamefont {S.~M.}\
  \bibnamefont {Girvin}}, \ and\ \bibinfo {author} {\bibfnamefont {R.~J.}\
  \bibnamefont {Schoelkopf}},\ }\bibfield  {title} {\enquote {\bibinfo {title}
  {Strong coupling of a single photon to a superconducting qubit using circuit
  quantum electrodynamics},}\ }\href@noop {} {\bibfield  {journal} {\bibinfo
  {journal} {Nature}\ }\textbf {\bibinfo {volume} {431}},\ \bibinfo {pages}
  {162--167} (\bibinfo {year} {2004})}\BibitemShut {NoStop}%
\bibitem [{\citenamefont {Chiorescu}\ \emph {et~al.}(2004)\citenamefont
  {Chiorescu}, \citenamefont {Bertet}, \citenamefont {Semba}, \citenamefont
  {Nakamura}, \citenamefont {Harmans},\ and\ \citenamefont
  {Mooij}}]{Chiorescu2004}%
  \BibitemOpen
  \bibfield  {author} {\bibinfo {author} {\bibfnamefont {I.}~\bibnamefont
  {Chiorescu}}, \bibinfo {author} {\bibfnamefont {P.}~\bibnamefont {Bertet}},
  \bibinfo {author} {\bibfnamefont {K.}~\bibnamefont {Semba}}, \bibinfo
  {author} {\bibfnamefont {Y.}~\bibnamefont {Nakamura}}, \bibinfo {author}
  {\bibfnamefont {C.J.P.M.}\ \bibnamefont {Harmans}}, \ and\ \bibinfo {author}
  {\bibfnamefont {J.~E.}\ \bibnamefont {Mooij}},\ }\bibfield  {title} {\enquote
  {\bibinfo {title} {Coherent dynamics of a flux qubit coupled to a harmonic
  oscillator},}\ }\href@noop {} {\bibfield  {journal} {\bibinfo  {journal}
  {Nature}\ }\textbf {\bibinfo {volume} {431}},\ \bibinfo {pages} {159--162}
  (\bibinfo {year} {2004})}\BibitemShut {NoStop}%
\bibitem [{\citenamefont {Niemczyk}\ \emph {et~al.}(2010)\citenamefont
  {Niemczyk}, \citenamefont {Deppe}, \citenamefont {Huebl}, \citenamefont
  {Menzel}, \citenamefont {Hocke}, \citenamefont {Schwarz}, \citenamefont
  {Garc{\'\i}a-Ripoll}, \citenamefont {Zueco}, \citenamefont {H{\"u}mmer},
  \citenamefont {Solano}, \citenamefont {Marx},\ and\ \citenamefont
  {Gross}}]{Niemczyk2010}%
  \BibitemOpen
  \bibfield  {author} {\bibinfo {author} {\bibfnamefont {T.}~\bibnamefont
  {Niemczyk}}, \bibinfo {author} {\bibfnamefont {F.}~\bibnamefont {Deppe}},
  \bibinfo {author} {\bibfnamefont {H.}~\bibnamefont {Huebl}}, \bibinfo
  {author} {\bibfnamefont {E.~P.}\ \bibnamefont {Menzel}}, \bibinfo {author}
  {\bibfnamefont {F.}~\bibnamefont {Hocke}}, \bibinfo {author} {\bibfnamefont
  {M.~J.}\ \bibnamefont {Schwarz}}, \bibinfo {author} {\bibfnamefont {J.~J.}\
  \bibnamefont {Garc{\'\i}a-Ripoll}}, \bibinfo {author} {\bibfnamefont
  {D.}~\bibnamefont {Zueco}}, \bibinfo {author} {\bibfnamefont
  {T.}~\bibnamefont {H{\"u}mmer}}, \bibinfo {author} {\bibfnamefont
  {E.}~\bibnamefont {Solano}}, \bibinfo {author} {\bibfnamefont
  {A.}~\bibnamefont {Marx}}, \ and\ \bibinfo {author} {\bibfnamefont
  {R.}~\bibnamefont {Gross}},\ }\bibfield  {title} {\enquote {\bibinfo {title}
  {Circuit quantum electrodynamics in the ultrastrong-coupling regime},}\
  }\href@noop {} {\bibfield  {journal} {\bibinfo  {journal} {Nature Phys.}\
  }\textbf {\bibinfo {volume} {6}},\ \bibinfo {pages} {772--776} (\bibinfo
  {year} {2010})}\BibitemShut {NoStop}%
\bibitem [{\citenamefont {Forn-D{\'\i}az}\ \emph {et~al.}(2010)\citenamefont
  {Forn-D{\'\i}az}, \citenamefont {Lisenfeld}, \citenamefont {Marcos},
  \citenamefont {Garc{\'\i}a-Ripoll}, \citenamefont {Solano}, \citenamefont
  {Harmans},\ and\ \citenamefont {Mooij}}]{Forn-Diaz2010}%
  \BibitemOpen
  \bibfield  {author} {\bibinfo {author} {\bibfnamefont {P.}~\bibnamefont
  {Forn-D{\'\i}az}}, \bibinfo {author} {\bibfnamefont {J.}~\bibnamefont
  {Lisenfeld}}, \bibinfo {author} {\bibfnamefont {D.}~\bibnamefont {Marcos}},
  \bibinfo {author} {\bibfnamefont {J.~J.}\ \bibnamefont {Garc{\'\i}a-Ripoll}},
  \bibinfo {author} {\bibfnamefont {E.}~\bibnamefont {Solano}}, \bibinfo
  {author} {\bibfnamefont {C.J.P.M.}\ \bibnamefont {Harmans}}, \ and\ \bibinfo
  {author} {\bibfnamefont {J.~E.}\ \bibnamefont {Mooij}},\ }\bibfield  {title}
  {\enquote {\bibinfo {title} {Observation of the {B}loch-{S}iegert shift in a
  qubit-oscillator system in the ultrastrong coupling regime},}\ }\href@noop {}
  {\bibfield  {journal} {\bibinfo  {journal} {Phys. Rev. Lett.}\ }\textbf
  {\bibinfo {volume} {105}},\ \bibinfo {pages} {237001} (\bibinfo {year}
  {2010})}\BibitemShut {NoStop}%
\bibitem [{\citenamefont {Fedorov}\ \emph {et~al.}(2010)\citenamefont
  {Fedorov}, \citenamefont {Feofanov}, \citenamefont {Macha}, \citenamefont
  {Forn-D{\'\i}az}, \citenamefont {Harmans},\ and\ \citenamefont
  {Mooij}}]{Fedorov2010}%
  \BibitemOpen
  \bibfield  {author} {\bibinfo {author} {\bibfnamefont {A.}~\bibnamefont
  {Fedorov}}, \bibinfo {author} {\bibfnamefont {A.~K.}\ \bibnamefont
  {Feofanov}}, \bibinfo {author} {\bibfnamefont {P.}~\bibnamefont {Macha}},
  \bibinfo {author} {\bibfnamefont {P.}~\bibnamefont {Forn-D{\'\i}az}},
  \bibinfo {author} {\bibfnamefont {C.~J.~P.~M.}\ \bibnamefont {Harmans}}, \
  and\ \bibinfo {author} {\bibfnamefont {J.~E.}\ \bibnamefont {Mooij}},\
  }\bibfield  {title} {\enquote {\bibinfo {title} {Strong coupling of a quantum
  oscillator to a flux qubit at its symmetry point},}\ }\href@noop {}
  {\bibfield  {journal} {\bibinfo  {journal} {Phys. Rev. Lett.}\ }\textbf
  {\bibinfo {volume} {105}},\ \bibinfo {pages} {060503} (\bibinfo {year}
  {2010})}\BibitemShut {NoStop}%
\bibitem [{\citenamefont {Forn-D{\'\i}az}\ \emph {et~al.}(2015)\citenamefont
  {Forn-D{\'\i}az}, \citenamefont {Romero}, \citenamefont {Harmans},
  \citenamefont {Solano},\ and\ \citenamefont {Mooij}}]{Forn-Diaz2015}%
  \BibitemOpen
  \bibfield  {author} {\bibinfo {author} {\bibfnamefont {P.}~\bibnamefont
  {Forn-D{\'\i}az}}, \bibinfo {author} {\bibfnamefont {G.}~\bibnamefont
  {Romero}}, \bibinfo {author} {\bibfnamefont {C.~J. P.~M.}\ \bibnamefont
  {Harmans}}, \bibinfo {author} {\bibfnamefont {E.}~\bibnamefont {Solano}}, \
  and\ \bibinfo {author} {\bibfnamefont {J.~E.}\ \bibnamefont {Mooij}},\
  }\bibfield  {title} {\enquote {\bibinfo {title} {Broken selection rule in the
  quantum {R}abi model},}\ }\href@noop {} {\bibfield  {journal} {\bibinfo
  {journal} {arXiv preprint arXiv:1510.03379}\ } (\bibinfo {year}
  {2015})}\BibitemShut {NoStop}%
\bibitem [{\citenamefont {De~Liberato}\ \emph {et~al.}(2009)\citenamefont
  {De~Liberato}, \citenamefont {Gerace}, \citenamefont {Carusotto},\ and\
  \citenamefont {Ciuti}}]{DeLiberato2009}%
  \BibitemOpen
  \bibfield  {author} {\bibinfo {author} {\bibfnamefont {S.}~\bibnamefont
  {De~Liberato}}, \bibinfo {author} {\bibfnamefont {D.}~\bibnamefont {Gerace}},
  \bibinfo {author} {\bibfnamefont {I.}~\bibnamefont {Carusotto}}, \ and\
  \bibinfo {author} {\bibfnamefont {C.}~\bibnamefont {Ciuti}},\ }\bibfield
  {title} {\enquote {\bibinfo {title} {Extracavity quantum vacuum radiation
  from a single qubit},}\ }\href@noop {} {\bibfield  {journal} {\bibinfo
  {journal} {Phys. Rev. A}\ }\textbf {\bibinfo {volume} {80}},\ \bibinfo
  {pages} {053810} (\bibinfo {year} {2009})}\BibitemShut {NoStop}%
\bibitem [{\citenamefont {Ai}\ \emph {et~al.}(2010)\citenamefont {Ai},
  \citenamefont {Li}, \citenamefont {Zheng}, \citenamefont {Sun} \emph
  {et~al.}}]{Ai2010}%
  \BibitemOpen
  \bibfield  {author} {\bibinfo {author} {\bibfnamefont {Q.}~\bibnamefont
  {Ai}}, \bibinfo {author} {\bibfnamefont {Y.}~\bibnamefont {Li}}, \bibinfo
  {author} {\bibfnamefont {H.}~\bibnamefont {Zheng}}, \bibinfo {author}
  {\bibfnamefont {C.~P.}\ \bibnamefont {Sun}},  \emph {et~al.},\ }\bibfield
  {title} {\enquote {\bibinfo {title} {Quantum anti-{Z}eno effect without
  rotating wave approximation},}\ }\href@noop {} {\bibfield  {journal}
  {\bibinfo  {journal} {Phys. Rev. A}\ }\textbf {\bibinfo {volume} {81}},\
  \bibinfo {pages} {042116} (\bibinfo {year} {2010})}\BibitemShut {NoStop}%
\bibitem [{\citenamefont {Cao}\ \emph {et~al.}(2010)\citenamefont {Cao},
  \citenamefont {You}, \citenamefont {Zheng}, \citenamefont {Kofman},\ and\
  \citenamefont {Nori}}]{Cao2010}%
  \BibitemOpen
  \bibfield  {author} {\bibinfo {author} {\bibfnamefont {X.}~\bibnamefont
  {Cao}}, \bibinfo {author} {\bibfnamefont {J.~Q.}\ \bibnamefont {You}},
  \bibinfo {author} {\bibfnamefont {H.}~\bibnamefont {Zheng}}, \bibinfo
  {author} {\bibfnamefont {A.~G.}\ \bibnamefont {Kofman}}, \ and\ \bibinfo
  {author} {\bibfnamefont {F.}~\bibnamefont {Nori}},\ }\bibfield  {title}
  {\enquote {\bibinfo {title} {Dynamics and quantum {Z}eno effect for a qubit
  in either a low-or high-frequency bath beyond the rotating-wave
  approximation},}\ }\href@noop {} {\bibfield  {journal} {\bibinfo  {journal}
  {Phys. Rev. A}\ }\textbf {\bibinfo {volume} {82}},\ \bibinfo {pages} {022119}
  (\bibinfo {year} {2010})}\BibitemShut {NoStop}%
\bibitem [{\citenamefont {Cao}\ \emph {et~al.}(2011)\citenamefont {Cao},
  \citenamefont {You}, \citenamefont {Zheng},\ and\ \citenamefont
  {Nori}}]{Cao2011}%
  \BibitemOpen
  \bibfield  {author} {\bibinfo {author} {\bibfnamefont {X.}~\bibnamefont
  {Cao}}, \bibinfo {author} {\bibfnamefont {J.~Q.}\ \bibnamefont {You}},
  \bibinfo {author} {\bibfnamefont {H.}~\bibnamefont {Zheng}}, \ and\ \bibinfo
  {author} {\bibfnamefont {F.}~\bibnamefont {Nori}},\ }\bibfield  {title}
  {\enquote {\bibinfo {title} {A qubit strongly coupled to a resonant cavity:
  asymmetry of the spontaneous emission spectrum beyond the rotating wave
  approximation},}\ }\href@noop {} {\bibfield  {journal} {\bibinfo  {journal}
  {New J. Phys.}\ }\textbf {\bibinfo {volume} {13}},\ \bibinfo {pages} {073002}
  (\bibinfo {year} {2011})}\BibitemShut {NoStop}%
\bibitem [{\citenamefont {Stassi}\ \emph {et~al.}(2013)\citenamefont {Stassi},
  \citenamefont {Ridolfo}, \citenamefont {Di~Stefano}, \citenamefont
  {Hartmann},\ and\ \citenamefont {Savasta}}]{Stassi2013}%
  \BibitemOpen
  \bibfield  {author} {\bibinfo {author} {\bibfnamefont {R.}~\bibnamefont
  {Stassi}}, \bibinfo {author} {\bibfnamefont {A.}~\bibnamefont {Ridolfo}},
  \bibinfo {author} {\bibfnamefont {O.}~\bibnamefont {Di~Stefano}}, \bibinfo
  {author} {\bibfnamefont {M.~J.}\ \bibnamefont {Hartmann}}, \ and\ \bibinfo
  {author} {\bibfnamefont {S.}~\bibnamefont {Savasta}},\ }\bibfield  {title}
  {\enquote {\bibinfo {title} {Spontaneous conversion from virtual to real
  photons in the ultrastrong-coupling regime},}\ }\href@noop {} {\bibfield
  {journal} {\bibinfo  {journal} {Phys. Rev. Lett.}\ }\textbf {\bibinfo
  {volume} {110}},\ \bibinfo {pages} {243601} (\bibinfo {year}
  {2013})}\BibitemShut {NoStop}%
\bibitem [{\citenamefont {Ridolfo}\ \emph {et~al.}(2013)\citenamefont
  {Ridolfo}, \citenamefont {Savasta},\ and\ \citenamefont
  {Hartmann}}]{Ridolfo2013}%
  \BibitemOpen
  \bibfield  {author} {\bibinfo {author} {\bibfnamefont {A.}~\bibnamefont
  {Ridolfo}}, \bibinfo {author} {\bibfnamefont {S.}~\bibnamefont {Savasta}}, \
  and\ \bibinfo {author} {\bibfnamefont {M.~J.}\ \bibnamefont {Hartmann}},\
  }\bibfield  {title} {\enquote {\bibinfo {title} {Nonclassical radiation from
  thermal cavities in the ultrastrong coupling regime},}\ }\href@noop {}
  {\bibfield  {journal} {\bibinfo  {journal} {Phys. Rev. Lett.}\ }\textbf
  {\bibinfo {volume} {110}},\ \bibinfo {pages} {163601} (\bibinfo {year}
  {2013})}\BibitemShut {NoStop}%
\bibitem [{\citenamefont {Garziano}\ \emph {et~al.}(2013)\citenamefont
  {Garziano}, \citenamefont {Ridolfo}, \citenamefont {Stassi}, \citenamefont
  {Di~Stefano},\ and\ \citenamefont {Savasta}}]{Garziano2013}%
  \BibitemOpen
  \bibfield  {author} {\bibinfo {author} {\bibfnamefont {L.}~\bibnamefont
  {Garziano}}, \bibinfo {author} {\bibfnamefont {A.}~\bibnamefont {Ridolfo}},
  \bibinfo {author} {\bibfnamefont {R.}~\bibnamefont {Stassi}}, \bibinfo
  {author} {\bibfnamefont {O.}~\bibnamefont {Di~Stefano}}, \ and\ \bibinfo
  {author} {\bibfnamefont {S.}~\bibnamefont {Savasta}},\ }\bibfield  {title}
  {\enquote {\bibinfo {title} {Switching on and off of ultrastrong light-matter
  interaction: Photon statistics of quantum vacuum radiation},}\ }\href
  {\doibase 10.1103/PhysRevA.88.063829} {\bibfield  {journal} {\bibinfo
  {journal} {Phys. Rev. A}\ }\textbf {\bibinfo {volume} {88}},\ \bibinfo
  {pages} {063829} (\bibinfo {year} {2013})}\BibitemShut {NoStop}%
\bibitem [{\citenamefont {Garziano}\ \emph {et~al.}(2014)\citenamefont
  {Garziano}, \citenamefont {Stassi}, \citenamefont {Ridolfo}, \citenamefont
  {Di~Stefano},\ and\ \citenamefont {Savasta}}]{Garziano2014}%
  \BibitemOpen
  \bibfield  {author} {\bibinfo {author} {\bibfnamefont {L.}~\bibnamefont
  {Garziano}}, \bibinfo {author} {\bibfnamefont {R.}~\bibnamefont {Stassi}},
  \bibinfo {author} {\bibfnamefont {A.}~\bibnamefont {Ridolfo}}, \bibinfo
  {author} {\bibfnamefont {O.}~\bibnamefont {Di~Stefano}}, \ and\ \bibinfo
  {author} {\bibfnamefont {S.}~\bibnamefont {Savasta}},\ }\bibfield  {title}
  {\enquote {\bibinfo {title} {Vacuum-induced symmetry breaking in a
  superconducting quantum circuit},}\ }\href@noop {} {\bibfield  {journal}
  {\bibinfo  {journal} {Phys. Rev. A}\ }\textbf {\bibinfo {volume} {90}},\
  \bibinfo {pages} {043817} (\bibinfo {year} {2014})}\BibitemShut {NoStop}%
\bibitem [{\citenamefont {Huang}\ and\ \citenamefont {Law}(2014)}]{Huang2014}%
  \BibitemOpen
  \bibfield  {author} {\bibinfo {author} {\bibfnamefont {J.-F.}\ \bibnamefont
  {Huang}}\ and\ \bibinfo {author} {\bibfnamefont {C.~K.}\ \bibnamefont
  {Law}},\ }\bibfield  {title} {\enquote {\bibinfo {title} {Photon emission via
  vacuum-dressed intermediate states under ultrastrong coupling},}\ }\href@noop
  {} {\bibfield  {journal} {\bibinfo  {journal} {Phys. Rev. A}\ }\textbf
  {\bibinfo {volume} {89}},\ \bibinfo {pages} {033827} (\bibinfo {year}
  {2014})}\BibitemShut {NoStop}%
\bibitem [{\citenamefont {Zhao}\ \emph {et~al.}(2015)\citenamefont {Zhao},
  \citenamefont {Liu}, \citenamefont {Liu},\ and\ \citenamefont
  {Nori}}]{Zhao2015}%
  \BibitemOpen
  \bibfield  {author} {\bibinfo {author} {\bibfnamefont {Y.-J.}\ \bibnamefont
  {Zhao}}, \bibinfo {author} {\bibfnamefont {Y.-L.}\ \bibnamefont {Liu}},
  \bibinfo {author} {\bibfnamefont {Y.-X.}\ \bibnamefont {Liu}}, \ and\
  \bibinfo {author} {\bibfnamefont {F.}~\bibnamefont {Nori}},\ }\bibfield
  {title} {\enquote {\bibinfo {title} {Generating nonclassical photon states
  via longitudinal couplings between superconducting qubits and microwave
  fields},}\ }\href@noop {} {\bibfield  {journal} {\bibinfo  {journal} {Phys.
  Rev. A}\ }\textbf {\bibinfo {volume} {91}},\ \bibinfo {pages} {053820}
  (\bibinfo {year} {2015})}\BibitemShut {NoStop}%
\bibitem [{\citenamefont {Zhu}\ \emph {et~al.}(2013)\citenamefont {Zhu},
  \citenamefont {Ferguson}, \citenamefont {Manucharyan},\ and\ \citenamefont
  {Koch}}]{Zhu2013}%
  \BibitemOpen
  \bibfield  {author} {\bibinfo {author} {\bibfnamefont {G.}~\bibnamefont
  {Zhu}}, \bibinfo {author} {\bibfnamefont {D.~G.}\ \bibnamefont {Ferguson}},
  \bibinfo {author} {\bibfnamefont {V.~E.}\ \bibnamefont {Manucharyan}}, \ and\
  \bibinfo {author} {\bibfnamefont {J.}~\bibnamefont {Koch}},\ }\bibfield
  {title} {\enquote {\bibinfo {title} {Circuit {Q}{E}{D} with fluxonium qubits:
  Theory of the dispersive regime},}\ }\href@noop {} {\bibfield  {journal}
  {\bibinfo  {journal} {Phys. Rev. B}\ }\textbf {\bibinfo {volume} {87}},\
  \bibinfo {pages} {024510} (\bibinfo {year} {2013})}\BibitemShut {NoStop}%
\bibitem [{\citenamefont {Ma}\ and\ \citenamefont {Law}(2015)}]{Law2015}%
  \BibitemOpen
  \bibfield  {author} {\bibinfo {author} {\bibfnamefont {K.~K.~W.}\
  \bibnamefont {Ma}}\ and\ \bibinfo {author} {\bibfnamefont {C.~K.}\
  \bibnamefont {Law}},\ }\bibfield  {title} {\enquote {\bibinfo {title}
  {Three-photon resonance and adiabatic passage in the large-detuning {R}abi
  model},}\ }\href {\doibase 10.1103/PhysRevA.92.023842} {\bibfield  {journal}
  {\bibinfo  {journal} {Phys. Rev. A}\ }\textbf {\bibinfo {volume} {92}},\
  \bibinfo {pages} {023842} (\bibinfo {year} {2015})}\BibitemShut {NoStop}%
\bibitem [{\citenamefont {Garziano}\ \emph {et~al.}(2015)\citenamefont
  {Garziano}, \citenamefont {Stassi}, \citenamefont {Macr{\`\i}}, \citenamefont
  {Kockum}, \citenamefont {Savasta},\ and\ \citenamefont
  {Nori}}]{Garziano2015}%
  \BibitemOpen
  \bibfield  {author} {\bibinfo {author} {\bibfnamefont {L.}~\bibnamefont
  {Garziano}}, \bibinfo {author} {\bibfnamefont {R.}~\bibnamefont {Stassi}},
  \bibinfo {author} {\bibfnamefont {V.}~\bibnamefont {Macr{\`\i}}}, \bibinfo
  {author} {\bibfnamefont {A.~F.}\ \bibnamefont {Kockum}}, \bibinfo {author}
  {\bibfnamefont {S.}~\bibnamefont {Savasta}}, \ and\ \bibinfo {author}
  {\bibfnamefont {F.}~\bibnamefont {Nori}},\ }\bibfield  {title} {\enquote
  {\bibinfo {title} {Multiphoton quantum {R}abi oscillations in ultrastrong
  cavity {Q}{E}{D}},}\ }\href@noop {} {\bibfield  {journal} {\bibinfo
  {journal} {Phys. Rev. A}\ }\textbf {\bibinfo {volume} {92}},\ \bibinfo
  {pages} {063830} (\bibinfo {year} {2015})}\BibitemShut {NoStop}%
\bibitem [{\citenamefont {Liu}\ \emph {et~al.}(2005)\citenamefont {Liu},
  \citenamefont {You}, \citenamefont {Wei}, \citenamefont {Sun},\ and\
  \citenamefont {Nori}}]{Liu2005}%
  \BibitemOpen
  \bibfield  {author} {\bibinfo {author} {\bibfnamefont {Y.~X.}\ \bibnamefont
  {Liu}}, \bibinfo {author} {\bibfnamefont {J.~Q.}\ \bibnamefont {You}},
  \bibinfo {author} {\bibfnamefont {L.~F.}\ \bibnamefont {Wei}}, \bibinfo
  {author} {\bibfnamefont {C.~P.}\ \bibnamefont {Sun}}, \ and\ \bibinfo
  {author} {\bibfnamefont {F.}~\bibnamefont {Nori}},\ }\bibfield  {title}
  {\enquote {\bibinfo {title} {Optical selection rules and phase-dependent
  adiabatic state control in a superconducting quantum circuit},}\ }\href
  {\doibase 10.1103/PhysRevLett.95.087001} {\bibfield  {journal} {\bibinfo
  {journal} {Phys. Rev. Lett.}\ }\textbf {\bibinfo {volume} {95}},\ \bibinfo
  {pages} {087001} (\bibinfo {year} {2005})}\BibitemShut {NoStop}%
\bibitem [{\citenamefont {{Deppe}}\ \emph {et~al.}(2008)\citenamefont
  {{Deppe}}, \citenamefont {{Mariantoni}}, \citenamefont {{Menzel}},
  \citenamefont {{Marx}}, \citenamefont {{Saito}}, \citenamefont
  {{Kakuyanagi}}, \citenamefont {{Tanaka}}, \citenamefont {{Meno}},
  \citenamefont {{Semba}}, \citenamefont {{Takayanagi}}, \citenamefont
  {{Solano}},\ and\ \citenamefont {{Gross}}}]{Gross2008}%
  \BibitemOpen
  \bibfield  {author} {\bibinfo {author} {\bibfnamefont {F.}~\bibnamefont
  {{Deppe}}}, \bibinfo {author} {\bibfnamefont {M.}~\bibnamefont
  {{Mariantoni}}}, \bibinfo {author} {\bibfnamefont {E.~P.}\ \bibnamefont
  {{Menzel}}}, \bibinfo {author} {\bibfnamefont {A.}~\bibnamefont {{Marx}}},
  \bibinfo {author} {\bibfnamefont {S.}~\bibnamefont {{Saito}}}, \bibinfo
  {author} {\bibfnamefont {K.}~\bibnamefont {{Kakuyanagi}}}, \bibinfo {author}
  {\bibfnamefont {H.}~\bibnamefont {{Tanaka}}}, \bibinfo {author}
  {\bibfnamefont {T.}~\bibnamefont {{Meno}}}, \bibinfo {author} {\bibfnamefont
  {K.}~\bibnamefont {{Semba}}}, \bibinfo {author} {\bibfnamefont
  {H.}~\bibnamefont {{Takayanagi}}}, \bibinfo {author} {\bibfnamefont
  {E.}~\bibnamefont {{Solano}}}, \ and\ \bibinfo {author} {\bibfnamefont
  {R.}~\bibnamefont {{Gross}}},\ }\bibfield  {title} {\enquote {\bibinfo
  {title} {{Two-photon probe of the Jaynes Cummings model and controlled
  symmetry breaking in circuit {Q}{E}{D}}},}\ }\href@noop {} {\bibfield
  {journal} {\bibinfo  {journal} {Nature Phys.}\ }\textbf {\bibinfo {volume}
  {4}},\ \bibinfo {pages} {686} (\bibinfo {year} {2008})}\BibitemShut {NoStop}%
\bibitem [{Sup()}]{SuppMat}%
  \BibitemOpen
  \href@noop {} {}\bibinfo {note} {See Supplemental Material for more
  details.}\BibitemShut {Stop}%
\bibitem [{\citenamefont {Ridolfo}\ \emph {et~al.}(2012)\citenamefont
  {Ridolfo}, \citenamefont {Leib}, \citenamefont {Savasta},\ and\ \citenamefont
  {Hartmann}}]{Ridolfo2012}%
  \BibitemOpen
  \bibfield  {author} {\bibinfo {author} {\bibfnamefont {A.}~\bibnamefont
  {Ridolfo}}, \bibinfo {author} {\bibfnamefont {M.}~\bibnamefont {Leib}},
  \bibinfo {author} {\bibfnamefont {S.}~\bibnamefont {Savasta}}, \ and\
  \bibinfo {author} {\bibfnamefont {M.~J.}\ \bibnamefont {Hartmann}},\
  }\bibfield  {title} {\enquote {\bibinfo {title} {Photon blockade in the
  ultrastrong coupling regime},}\ }\href@noop {} {\bibfield  {journal}
  {\bibinfo  {journal} {Phys. Rev. Lett.}\ }\textbf {\bibinfo {volume} {109}},\
  \bibinfo {pages} {193602} (\bibinfo {year} {2012})}\BibitemShut {NoStop}%
\bibitem [{\citenamefont {Hoffman}\ \emph {et~al.}(2011)\citenamefont
  {Hoffman}, \citenamefont {Srinivasan}, \citenamefont {Schmidt}, \citenamefont
  {Spietz}, \citenamefont {Aumentado}, \citenamefont {T{\"u}reci},\ and\
  \citenamefont {Houck}}]{Hoffman2011}%
  \BibitemOpen
  \bibfield  {author} {\bibinfo {author} {\bibfnamefont {A.~J.}\ \bibnamefont
  {Hoffman}}, \bibinfo {author} {\bibfnamefont {S.~J.}\ \bibnamefont
  {Srinivasan}}, \bibinfo {author} {\bibfnamefont {S.}~\bibnamefont {Schmidt}},
  \bibinfo {author} {\bibfnamefont {L.}~\bibnamefont {Spietz}}, \bibinfo
  {author} {\bibfnamefont {J.}~\bibnamefont {Aumentado}}, \bibinfo {author}
  {\bibfnamefont {H.~E.}\ \bibnamefont {T{\"u}reci}}, \ and\ \bibinfo {author}
  {\bibfnamefont {A.~A.}\ \bibnamefont {Houck}},\ }\bibfield  {title} {\enquote
  {\bibinfo {title} {Dispersive photon blockade in a superconducting
  circuit},}\ }\href@noop {} {\bibfield  {journal} {\bibinfo  {journal} {Phys.
  Rev. Lett.}\ }\textbf {\bibinfo {volume} {107}},\ \bibinfo {pages} {053602}
  (\bibinfo {year} {2011})}\BibitemShut {NoStop}%
\bibitem [{\citenamefont {Majer}\ \emph {et~al.}(2007)\citenamefont {Majer},
  \citenamefont {Chow}, \citenamefont {Gambetta}, \citenamefont {Koch},
  \citenamefont {Johnson}, \citenamefont {Schreier}, \citenamefont {Frunzio},
  \citenamefont {Schuster}, \citenamefont {Houck}, \citenamefont {Wallraff}
  \emph {et~al.}}]{Majer2007}%
  \BibitemOpen
  \bibfield  {author} {\bibinfo {author} {\bibfnamefont {J.}~\bibnamefont
  {Majer}}, \bibinfo {author} {\bibfnamefont {J.M.}\ \bibnamefont {Chow}},
  \bibinfo {author} {\bibfnamefont {J.M.}\ \bibnamefont {Gambetta}}, \bibinfo
  {author} {\bibfnamefont {J.}~\bibnamefont {Koch}}, \bibinfo {author}
  {\bibfnamefont {B.R.}\ \bibnamefont {Johnson}}, \bibinfo {author}
  {\bibfnamefont {J.A.}\ \bibnamefont {Schreier}}, \bibinfo {author}
  {\bibfnamefont {L.}~\bibnamefont {Frunzio}}, \bibinfo {author} {\bibfnamefont
  {D.I.}\ \bibnamefont {Schuster}}, \bibinfo {author} {\bibfnamefont {A.A.}\
  \bibnamefont {Houck}}, \bibinfo {author} {\bibfnamefont {A.}~\bibnamefont
  {Wallraff}},  \emph {et~al.},\ }\bibfield  {title} {\enquote {\bibinfo
  {title} {Coupling superconducting qubits via a cavity bus},}\ }\href@noop {}
  {\bibfield  {journal} {\bibinfo  {journal} {Nature}\ }\textbf {\bibinfo
  {volume} {449}},\ \bibinfo {pages} {443} (\bibinfo {year}
  {2007})}\BibitemShut {NoStop}%
\bibitem [{\citenamefont {Nation}\ \emph {et~al.}(2012)\citenamefont {Nation},
  \citenamefont {Johansson}, \citenamefont {Blencowe},\ and\ \citenamefont
  {Nori}}]{Norirmp}%
  \BibitemOpen
  \bibfield  {author} {\bibinfo {author} {\bibfnamefont {P.~D.}\ \bibnamefont
  {Nation}}, \bibinfo {author} {\bibfnamefont {J.~R.}\ \bibnamefont
  {Johansson}}, \bibinfo {author} {\bibfnamefont {M.~P.}\ \bibnamefont
  {Blencowe}}, \ and\ \bibinfo {author} {\bibfnamefont {F.}~\bibnamefont
  {Nori}},\ }\bibfield  {title} {\enquote {\bibinfo {title} {Stimulating
  uncertainty: Amplifying the quantum vacuum with superconducting circuits},}\
  }\href {\doibase 10.1103/RevModPhys.84.1} {\bibfield  {journal} {\bibinfo
  {journal} {Rev. Mod. Phys.}\ }\textbf {\bibinfo {volume} {84}},\ \bibinfo
  {pages} {1--24} (\bibinfo {year} {2012})}\BibitemShut {NoStop}%
\end{thebibliography}%

\end{document}